\documentclass[12pt]{article}
\usepackage{epsf,epsfig,float,amssymb,latexsym,amsmath,amsthm,fancyhdr}
\usepackage{graphics,psfrag,longtable}
\newcommand{\vp}{\mathbf{p}}
\newcommand{\bl}{\begin{aligned}}
\newcommand{\el}{\end{aligned}}

\newcommand{\la}{\langle}
\newcommand{\ra}{\rangle}

\newcommand{\be}{\begin{equation}}
\newcommand{\ee}{\end{equation}}
\newcommand{\ba}{\begin{eqnarray}}
\newcommand{\ea}{\end{eqnarray}}
\newcommand{\bg}{\begin{align}}
\newcommand{\egg}{\end{align}}

\newcommand{\nn}{\nonumber}

\newcommand{\krig}[1]{\stackrel{\circ}{#1}}
\newcommand{\barr}[1]{\not\mathrel  #1}

\begin{document}

\thispagestyle{empty}

\vspace{2cm}

\begin{center}
{\Large{\bf  $\pi N$ scattering in relativistic baryon chiral perturbation theory revisited}}
\end{center}
\vspace{.5cm}

\begin{center}
{\large J.~M.~Alarc\'on$^1$, J.~Mart\'{\i}n~Camalich$^{2,3}$, J.~A. Oller$^1$ and L.~Alvarez-Ruso$^{2,4}$}
\end{center}

\begin{center}
$^1${\it {\it Departamento de F\'{\i}sica. Universidad de Murcia. E-30071,
Murcia. Spain}}\\
$^2${\it {\it Departamento de F\'{\i}sica Te\'orica and IFIC, Universidad de Valencia-CSIC, E-46071 Spain}}\\
$^3${\it {\it Department of Physics and Astronomy, University of Sussex, BN1 9QH, Brighton, UK}}\\
$^4${\it {\it Centro de F\'{\i}sica Computacional. Departamento de F\'{\i}sica. Universidade de Coimbra, Portugal}}
\end{center}
\vspace{1cm}

\begin{abstract}
\noindent
We have analyzed pion-nucleon scattering using the manifestly relativistic covariant framework of 
Infrared Regularization up to ${\cal O}(q^3)$ in the chiral expansion, where $q$ is a generic small momentum.
 We describe the low-energy phase shifts with a similar quality as previously achieved with Heavy Baryon Chiral Perturbation 
Theory, $\sqrt{s}\lesssim1.14$~GeV. New values are provided for the ${\cal O}(q^2)$ and  ${\cal O}(q^3)$ 
low-energy constants, which are compared with previous determinations. This is also the case for the scattering lengths and volumes. 
Finally, we have unitarized the previous amplitudes  and as a result 
the energy range where data are reproduced increases significantly. 
  \end{abstract}

\vspace{2cm}


\newpage


\section{Introduction}
\label{sec:intro}
\def\theequation{\arabic{section}.\arabic{equation}}
\setcounter{equation}{0}
Pion-nucleon scattering is a fundamental process involving the lightest meson and baryon. Therefore, at low energies it is  amenable to be studied by the 
low-energy effective field theory of QCD, Chiral Perturbation Theory (CHPT) \cite{weinberg,gasser1}, which takes into account both the spontaneous as well as the explicit breaking of chiral symmetry in strong interactions. While the baryon field transforms linearly under the chiral group the pions do non-linearly  \cite{cole}. A first step in extending CHPT to the systems with baryon number one was undertaken in ref.~\cite{gasser2}. Contrary to standard CHPT, it was established that due to the presence of the large nucleon mass, loops do not respect the chiral power counting, and the lower order counterterms are renormalized because of higher order loops. The power counting was recovered by applying Heavy Baryon CHPT (HBCHPT) \cite{jenkins}, where the heavy components of the baryon fields are integrated out \cite{kambor,review} so that  manifest Lorentz invariance 
is lost. On the other hand,  for some loop functions  the expansion in inverse powers of the nucleon mass and the loop integration  do not commute  so that the non-relativistic expansion does not converge \cite{becher,review}. The recovery of the power counting, while keeping manifest Lorentz invariance,
 was  achieved by the Infrared Regularization method (IR) \cite{becher}, based on the 
 ideas in ref.~\cite{elli}. 
IR was extended to the multi-nucleon sector \cite{goity} and to multi-loop diagrams \cite{gegen}.  
Another relativistic approach to baryon CHPT is the so-called extended-on-mass-shell (EOMS) renormalization scheme \cite{eoms1,eoms2}. The latter is based on removing explicitly  the power counting breaking terms appearing in the loop
 integrals in dimensional regularization since they are re-absorbed by the finite set of low-energy counterterms up to the order the calculation is performed. For a  recent review on  baryon CHPT see ref.~\cite{bernard}.

Here we focus on the application of IR CHPT methods to low-energy pion-nucleon scattering. In HBCHPT there is already an extensive list of detailed calculations with several degrees of precision. 
In refs.~\cite{mojzis,elli,fettes3} an ${\cal O}(q^3)$ calculation is performed, with the additional inclusion of  the $\Delta(1232)$ in ref.~\cite{elli}. 
The calculation of pion-nucleon scattering was  extended up to ${\cal O}(q^4)$ in ref.~\cite{fettes4}, while isospin violation (including both strong isospin breaking terms and electromagnetism) is worked out up to ${\cal O}(q^3)$ in ref.~\cite{fettes_ep}.  The same authors also studied the 
influence of the $\Delta$-isobar within the small $\epsilon$-expansion \cite{hemmert}  up to ${\cal O}(\epsilon^3)$ in ref.~\cite{fettes_small}.\footnote{Other chiral power-countings including the explicit $\Delta(1232)$ resonance are the $\delta$-expansion \cite{pascalutsa12} and the more recent  one of ref.~\cite{kolck}.} 
 Isospin breaking corrections for the pion-nucleon scattering lengths to ${\cal O}(q^3)$ 
are calculated  within IR in refs.~\cite{lipartia,hoferichter}. This is part of an on-going effort for providing high precision determinations 
of the pion-nucleon scattering lengths (a recent review on this issue is ref.~\cite{gassrev}).  
Within IR CHPT $\pi N$ scattering was already considered in refs.~\cite{beche2,elli2}. Ref.~\cite{beche2} performed an ${\cal O}(q^4)$ one-loop calculation. 
 Its main conclusion was that the one-loop representation is not precise enough to allow a sufficiently accurate extrapolation of physical data to the Cheng-Dashen point.
 On the other hand, ref.~\cite{elli2} was interested in the complementary aspects  of comparing IR CHPT at ${\cal O}(q^3)$ to data and to previous 
HBCHPT studies \cite{elli,fettes3,fettes4}.
 The conclusions were  unexpected and rather pessimistic. The description obtained  was restricted to  very low center-of-mass (CM) pion kinetic energy (less than 40~MeV),  such that the IR results badly diverge from the experimental values above that energy in several partial waves \cite{elli2}. In comparison, the resulting phase shifts in HBCHPT \cite{elli,fettes3}  fit pion-nucleon phase shifts up to significantly higher energies and then start deviating smoothly from data.   Last but not least, ref.~\cite{elli2} also found 
 an unrealistically large  violation (20--30\%) of the Goldberger-Treiman (GT) relation  \cite{goldberger} for the pion-nucleon coupling. 
 As we shall show, our results are somewhat more optimistic that those of ref.~\cite{elli2} because we obtain that IR is able to describe low-energy pion-nucleon scattering comparable to HBCHPT at ${\cal O}(q^3)$. Nevertheless, the caveat about the large violation of the GT 
relation in a full IR calculation at ${\cal O}(q^3)$ remains. When this calculation is restricted to strict ${\cal O}(q^3)$, as in HBCHPT,  
more realistic values around 2\% are obtained for the violation of the GT relation. 

As a consequence of unitarity,  $\pi N$ partial wave amplitudes develop a right-hand  or unitarity cut with a branch 
point at the reaction threshold. The first derivative of the partial waves at this point is singular. 
Based on this, ref.~\cite{beche2} advocates for applying the chiral expansion to the subthreshold region of the 
 $\pi N$ scattering amplitude  where it is expected to be  smoother. 
 This singularity can also be avoided by applying the chiral expansion to an interaction kernel which, by construction, has no right-hand cut. 
This is the so-called Unitary CHPT (UCHPT) \cite{npa,nd,pin,plb}.
 One of the consequences of this framework,  is that the calculated $\pi N$ partial waves fulfill 
unitarity. We  compare in this work the purely perturbative results with those obtained by UCHPT, with the latter being able to 
fit data closely  up to higher energies as shown below.  There are other methods already employed to provide unitarized $\pi N$ amplitudes from the given chiral expansions, 
e.g.\cite{gomez,grana,gaspa}. We will also 
explore the region of the $\Delta(1232)$ resonance by including a Castillejo-Dalitz-Dyson (CDD) pole \cite{cdd} in the inverse of the
 amplitude. The resulting amplitude has the same discontinuities along the right- and left-hand-cuts as the one without 
the CDD poles. Traditionally, this was a serious drawback for the bootstrap hypothesis \cite{demo,demo2} 
since the source for dynamics  in this approach is given precisely by the discontinuities of the amplitudes along the cuts. 
From our present knowledge based on QCD the presence of these extra solutions (including an arbitrary number of CDD poles) 
can be expected on intuitive grounds since they would be required in order to accommodate pre-existing resonances due to the elementary degrees of freedom of  QCD. 
 Alternatively, 
one could also include explicitly within the effective field theory the $\Delta(1232)$ resonances 
as a massive field \cite{38,bora,hemmert,elli2,pascalutsa12,kolck}. 

After this introduction we give in section \ref{sec2} our conventions and Lagrangians employed, 
including kinematics and equations used to project into partial waves.  
The calculation at the one-loop level up to ${\cal O}(q^3)$ is performed in section \ref{sec3},
 where we compare directly with the expressions given in ref.~\cite{beche2}. We also present fits to the experimental data at the perturbative level and discuss the resulting values for the 
chiral counterterms, scattering lengths and volumes and the violation of the GT relation.
The resummation of the right-hand cut by means of UCHPT is undertaken in section \ref{sec4}, 
where we discuss the comparison with experimental data and the significant increase of the energy range  
for the reproduction of data.   The conclusions are given in section \ref{sec5}.

\section{Prelude: Generalities, kinematics and Lagrangians}
\label{sec2}
\def\theequation{\arabic{section}.\arabic{equation}}
\setcounter{equation}{0}

We consider the process $\pi^a(q) N(p,\sigma;\alpha)\to \pi^{a'}(q') N(p',\sigma';\alpha')$. 
Here $a$ and $a'$ denote the Cartesian coordinates in the isospin space of the initial and final pions with 
four-momentum $q$ and $q'$, respectively. 
Regarding the nucleons, $\sigma$($\sigma'$) and $ \alpha(\alpha')$ correspond to the third-components 
of spin and isospin of the initial (final) states, in order. The usual Mandelstam variables are 
defined as  $s=(p+q)^2=(p'+q')^2$, $t=(q-q')^2=(p-p')^2$ 
and $u=(p-q')^2=(p'-q)^2$, that fulfill $s+t+u=2M_\pi^2+2 m^2$ for on-shell scattering, with $m$ and $M_\pi$ the 
nucleon and pion mass, respectively. 
It is convenient to consider Lorentz- and isospin-invariant amplitudes (along our study  exact isospin symmetry is presummed.) 
 We then decompose the scattering amplitude as \cite{hohler}
\begin{align}
T_{ a a' }&=\delta_{a'a} T^++\frac{1}{2}[\tau_a,\tau_{a'}]T^{-}~,\nn\\
T^{\pm}&=\bar{u}(p',\sigma')\left[A^{\pm}+\frac{1}{2}(\barr{q}+{\barr{q}}\,')B^{\pm}\right]u(p,\sigma)~.
\label{apmbpmdef}
\end{align}
Here, the Pauli matrices are indicated by $ \tau_c$. 
In the next section we will  proceed with the calculation of $A^{\pm}$ and $B^ {\pm}$ 
perturbatively up to ${\cal O}(q^3)$. 

In IR  the Feynman diagrams for $\pi N$ scattering follow the 
standard chiral power counting \cite{wein}
\begin{align}
\nu=1+2L+\sum_{i}V_i(d_i+\frac{1}{2}n_i-2)~,
\label{counting}
\end{align}
where $L$ is the number of loops, $V_i$ is the number of vertices of type $i$ 
consisting of $n_i$ baryon fields (in our case $n_i=0$, 2) and $d_i$ pion derivatives or masses. 
In this way, a given Feynman diagram for $\pi N$ scattering counts as $q^\nu$. 
 For the calculation of pion-nucleon scattering up to ${\cal O}(q^3)$ we employ the chiral Lagrangian 
\begin{align}
{\cal L}_{CHPT}&={\cal L}_{\pi\pi}^{(2)}+
{\cal L}_{\pi\pi}^{(4)}+
{\cal L}_{\pi N}^{(1)}+
{\cal L}_{\pi N}^{(2)}+
{\cal L}_{\pi N}^{(3)}~,
\end{align}
where the superscript indicates the chiral order, according to eq.~\eqref{counting}. 
Here, ${\cal L}_{\pi\pi}^{(n)}$ refers to the purely mesonic Lagrangian without baryons and 
${\cal L}_{\pi N}^{(n)}$ corresponds to the one bilinear in the baryon fields. We follow the same notation 
as in ref.~\cite{beche2} to make easier the comparison. Then, 
\begin{align}
{\cal L}_{\pi\pi}^{(2)}&=\frac{F^2}{4}\la u_\mu u^\mu+\chi_+\ra ~\nn~,\\ 
{\cal L}_{\pi\pi}^{(4)}&=\frac{1}{16}\ell_4 \left(2 \la u_\mu u^\mu\ra \la \chi_+ \ra
+\la \chi_+\ra^2\right)+\ldots
\label{lagpi}
\end{align} 
where the ellipsis indicate terms that are not needed in the calculations given here. 
For the different symbols, $F$ is the pion weak decay constant in the chiral limit and 
\begin{align}
u^2&=U~,~u_\mu=i u^\dagger \partial_\mu U\, u^\dagger~,~\chi_{\pm}=u^\dagger \chi u^\dagger\pm u \chi^\dagger u~.
\end{align}
The explicit chiral symmetry breaking due to the non-vanishing quark masses (in the isospin limit $m_u=m_d=\hat{m}$) 
 is introduced through $\chi=2 B_0 \hat{m}$. The constant $B_0$ is proportional to the quark condensate in  the chiral limit 
$\la 0|\bar{q}^j q^i|0\ra=-B_0 F^2 \delta^{ij}$.
In the following we employ the so-called sigma-parameterization where
\begin{align}
U(x)&=\sqrt{1-\frac{\vec{\pi}(x)^2}{F^2}}+i\frac{\vec{\pi}(x)\cdot \vec{\tau}}{F}
\end{align}
In eq.~\eqref{lagpi} we denote by $\la \cdots \ra$ the trace of the resulting $2\times 2$ matrix.
 For the pion-nucleon Lagrangian we have
\begin{align}
{\cal L}_{\pi N}^{(1)}&=\bar{\psi}(i\barr{D}-\krig{m})\psi+\frac{g}{2}\bar{\psi}\barr{u}\gamma_5 \psi~,\nn\\
{\cal L}_{\pi N}^{(2)}&=c_1 \la \chi_+\ra \bar{\psi}\psi-\frac{c_2}{4m^2}\la u_\mu u_\nu\ra(\bar{\psi}D^\mu D^\nu \psi+\hbox{h.c.})+\frac{c_3}{2}\la u_\mu u^\mu\ra \bar{\psi}\psi-\frac{c_4}{4}\bar{\psi}\gamma^\mu\gamma^\nu[u_\mu,u_\nu]\psi+\ldots~,\nn\\
{\cal L}_{\pi N}^{(3)}&=\bar{\psi}\Biggl(-\frac{d_1+d_2}{4m}([u_\mu,[D_\nu,u^\mu]+[D^\mu,u_\nu]]D^\nu 
+\hbox{h.c.})\nn\\
&+\frac{d_3}{12 m^3}([u_\mu,[D_\nu,u_\lambda]](D^\mu D^\nu D^\lambda+\hbox{sym.})+\hbox{h.c.})
+i\frac{d_5}{2 m}([\chi_-,u_\mu]D^\mu+\hbox{h.c.})\nn\\
&+i\frac{d_{14}-d_{15}}{8 m}\left(\sigma^{\mu \nu}\la [D_\lambda,u_\mu]u_\nu-u_\mu [D_\nu,u_\lambda]\ra 
D^\lambda+\hbox{h.c.}\right)\nn\\
&+\frac{d_{16}}{2}\gamma^\mu\gamma_5\la\chi_+\ra u_\mu+\frac{id_{18}}{2}\gamma^\mu \gamma_5 [D_\mu,\chi_-]\Biggr) \psi
+\ldots
\label{lagN}
\end{align}
In the previous equation $\krig{m}$ is the nucleon mass in the chiral limit ($m_u=m_d=0$) and the covariant derivative $D_\mu$ acting on the baryon fields is given by $\partial_\mu+\Gamma_\mu$ with $\Gamma_\mu= [u^\dagger,\partial_\mu u]/2$. 
The low-energy counterterms (LECs) $c_i$ and $d_i$ are not fixed by chiral symmetry and we fit them to $\pi N$ scattering data.  
Again only the  terms needed for the present study are shown in eq.~\eqref{lagN}. For more details on the definition and derivation of the different 
monomials  we refer to refs.~\cite{fettes3,opv}.

The free one-particle states are normalized according to the Lorentz-invariant normalization
\begin{align}
\la \vp',\sigma';\gamma|\vp,\sigma;\gamma\ra=
  2 E_p (2\pi)^3\delta(\vp'-\vp) \delta_{\sigma\sigma'}\delta_{\gamma\gamma'}~,
\end{align}
where $E_p$ is the energy of the particle with three-momentum $\vp$ and $\gamma$ indicates any internal quantum number. A free two-particle state 
is normalized accordingly and it can be decomposed in states with  well defined total spin $S$ and total angular momentum $J$. For $\pi N$ scattering $S=1/2$ and one has in the CM frame
\begin{align}
|\pi(-\vp;a)N(\vp,\sigma;\alpha)\ra&=\sqrt{4\pi}\sum_{\ell,m} (m \sigma \mu|\ell \frac{1}{2} J) Y_\ell^m(\hat{\mathbf{p}})^*|J \mu \ell;a \alpha \ra~,
\label{waves}
\end{align}
with $\hat{\vp}$ the unit vector  of the CM nucleon three-momentum $\vp$, $\ell$ the orbital angular momentum, $m$ its third component  and $ \mu=m+\sigma$ the third-component of the total angular momentum. 
 The Clebsch-Gordan coefficient is denoted by $(m_1 m_2 m_3|j_1 j_2 j_3)$, corresponding to the composition of the spins $j_1$ and $j_2$ (with third-components $m_1$ and $m_2$, in order) to give the third spin $j_3$, with third-component $m_3$. 
The state with total angular momentum well-defined, $|J \mu \ell;a \alpha\ra$, satisfies the normalization 
 condition 
\begin{align}
\la J' \mu' \ell';a' \alpha'|J \mu \ell;a \alpha\ra=\delta_{J J'}\delta_{\mu'\mu}\delta_{\ell \ell'}
\frac{4\pi \sqrt{s}}{|\vp|} \delta_{a'a}\delta_{\alpha'\alpha}~.
\label{jdef.norma}
\end{align}
The partial wave expansion of the $\pi N$ scattering amplitude can be worked out straightforwardly from eq.~\eqref{waves}. 
By definition, the initial baryon three-momentum $\vp$ gives the positive direction of the ${\mathbf{z}}$-axis. Inserting the series of eq.~\eqref{waves} one has for the scattering amplitude
\begin{align}
\la \pi(-\vp';a')N(\vp',\sigma';\alpha')|T|\pi(-\vp;a)N(\vp,\sigma;\alpha)\ra&=
4\pi\sum_{\ell,m,J}Y_\ell^0(\hat{\mathbf{z}})(m\sigma'\sigma|\ell\frac{1}{2}J)
(0\sigma\sigma|\ell\frac{1}{2}J) Y_{\ell}^m(\hat{\vp}') T_{J\ell}(s)~,
\label{series_t}
\end{align}
where $T$ is the T-matrix operator and $T_{J\ell}$ is the partial wave amplitude with total angular momentum $J$ and orbital angular momentum $\ell$. Notice that in eq.~\eqref{series_t} we   made use of the fact that $Y_\ell^m(\hat{\mathbf{z}})$ is non-zero only for $m=0$. Recall also that because of parity conservation partial wave amplitudes with different orbital angular momentum do not mix.
 From eq.~\eqref{series_t} it is straightforward to isolate $T_{J\ell}$ with the result
\begin{align}
T_{J\ell}(a',\alpha';a,\alpha)&=\frac{1}{\sqrt{4\pi(2\ell+1)}(0\sigma\sigma|\ell\frac{1}{2}J)}
\sum_{m,\sigma'}\int d\hat{\vp}'\,\la \pi(-\vp';a')N(\vp',\sigma';\alpha')|T|\pi(-\vp;a)N(\vp,\sigma;\alpha)\ra
\nn\\
&\times (m\sigma'\sigma|\ell\frac{1}{2}L) Y_\ell^m(\hat{\vp}')^*~.
\label{tjl}
\end{align}   
In the previous expression the  resulting $T_{J\ell}$ is of course independent 
of choice of $\sigma$. 

The relation between the  Cartesian and charge bases is given by 
\begin{align}
|\pi^+\ra&=\frac{1}{\sqrt{2}}(|\pi^1\ra+i|\pi^2\ra)~,\nn\\
|\pi^-\ra&=\frac{1}{\sqrt{2}}(|\pi^1\ra-i|\pi^2\ra)~,\nn\\
|\pi^0\ra&=|\pi^3\ra~.
\label{pion_charged}
\end{align}
 According to the previous definition of states 
$|\pi^+\ra=-|1,+1\ra$, $|\pi^-\ra=|1,-1\ra$ and $|\pi^0\ra=|\pi^3\ra=|1,0\ra$, where the states of the isospin basis 
are placed to the right of the equal sign. Notice the minus sign in the relationship for $|\pi^+\ra$.   Then, the amplitudes with  well-defined isospin, 
$I=3/2$ or 1/2, are  denoted by $T_{IJ\ell}$ and can be obtained employing the appropriate linear 
combinations of $T_ {J\ell}(a',\alpha';a,\alpha)$, eq.~\eqref{tjl}, in terms of standard  Clebsch-Gordan 
coefficients.

Due to the normalization of the states with well-defined total angular momentum, eq.~\eqref{jdef.norma}, 
the partial waves resulting from eq.~\eqref{tjl}  with 
well defined isospin satisfy the unitarity relation
\begin{align}
\hbox{Im}T_{IJ\ell}=\frac{|\vp|}{8\pi \sqrt{s}}|T_{IJ\ell}|^2
\label{unita}
\end{align}
for $|\vp|>0$ and below the inelastic threshold due the one-pion production at $|\vp|\simeq 210$~MeV. 
 Given the previous equation, the $S$-matrix element with well defined $I$, $J$ and $\ell$, denoted by
 $S_{I J\ell}$, corresponds to 
\begin{align}
S_{I J\ell}=1+i\frac{|\vp|}{4\pi\sqrt{s}}T_{I J\ell}~,
\label{s.def}
\end{align}
satisfying $S_{I J \ell}S_{I J \ell}^*=1$ in the elastic physical region. 
 In the same region we can then write 
\begin{align}
S_{I J\ell}=e^{2 i\delta_{I J\ell}}~,
\label{s.def.2}
\end{align}
with $\delta_{I J\ell}$ the corresponding phase shifts.

\section{Perturbative calculation and its results}
\label{sec3}
\def\theequation{\arabic{section}.\arabic{equation}}
\setcounter{equation}{0}

\begin{figure}[H]
\centerline{\epsfig{file=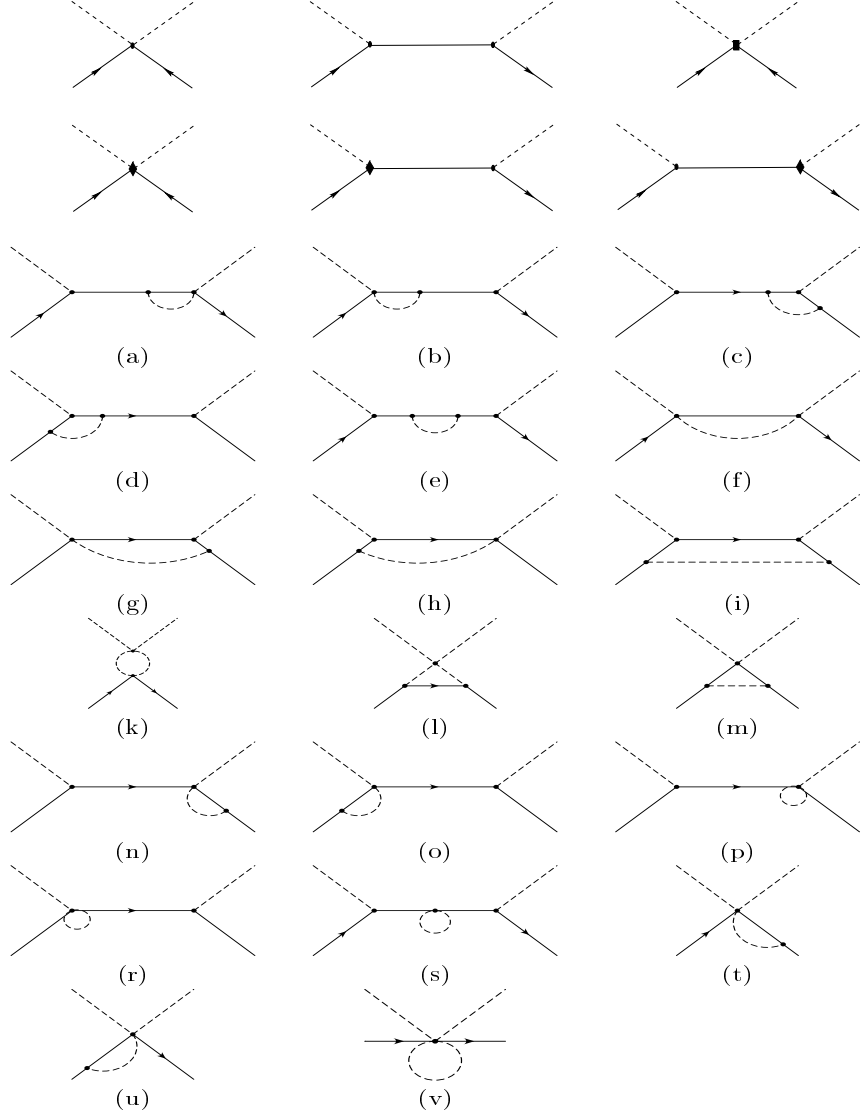,width=.65\textwidth,angle=0}}
\vspace{0.2cm}
\caption[pilf]{\protect \small Set of diagrams for $\pi N$ scattering up to-and-including ${\cal O}(q^3)$. 
The ${\cal O}(q)$ diagrams are the first two  in the first line, from left to right. 
The ${\cal O}(q^2)$ contributions 
correspond to the third one still in the first line, where the $d_i=2$, $n_2=2$ vertex is indicated 
with a square. The rest of the diagrams from the second line until the bottom of the figure
are ${\cal O}(q^3)$. The $d_i=3$, $n_i=2$ vertices are indicated with a diamond. The one-loop diagrams 
only have lowest order pion-nucleon vertices. 
 \label{fig:diag_list}}
\end{figure}

From eq.~\eqref{counting}  the leading order contribution to $\pi N$-scattering has $\nu=1$ and it 
consists only of the lowest order pion-nucleon vertices with $d_i=1$ and of no loops ($L=0$). 
These diagrams correspond to the first two topologies  shown from left to right in the first line of 
 fig.~\ref{fig:diag_list}, where all the diagrams up-to-and-including ${\cal O}(q^ 3)$  are shown.
The ${\cal O}(q^2)$ or next-to-leading order (NLO) 
contribution  still has no loops ($L=0$) and contains an ${\cal O}(q^2)$ vertex with $d_i=2$. It is shown by the
 third diagram in the first line of fig.~\ref{fig:diag_list}. The NLO pion-nucleon vertex is depicted by 
the filled square. The ${\cal O}(q^3)$ or next-to-next-to-leading (N$^2$LO) contributions consists of tree-level diagrams 
 with at least one vertex of $d_i=3$ type, with the other ones with $d_i=1$. They are shown 
by the diagrams in the second line of fig.~\ref{fig:diag_list}, where the diamond corresponds to the $d_i=3$ vertex.  
 Finally, at N$^2$LO one also has the one loop ($L=1$) diagrams involving only 
 vertices with $d_i=1$ from the LO pion-nucleon Lagrangian, ${\cal L}_{\pi N}^{(1)}$, and 
with $d_i=2$ from the LO pure mesonic Lagrangian, ${\cal L}_{\pi\pi}^{(2)}$. The one-loop diagrams are the rest of those shown in the figure and 
 are labelled with a latin letter (a)--(v). In addition, one also has the wave function renormalization of pions and nucleons affecting the LO contribution. 
The calculation is finally given in  terms of $m$, $F_\pi$ and $g_A$, which implies
some reshuffling of pieces once the constants $\krig{m}$, $F$ and $g$ in the chiral limit 
are expressed in terms of the physical $m$, $F_\pi$ and $g_A$ making use of their expressions at ${\cal O}(q^3)$ \cite{beche2}. In this work 
we employ the numerical values $F_\pi=92.4$~MeV, $M_\pi=139$~MeV, $m_N=939$~MeV, $g_A=1.267$ and $\mu=m_N$.

The set of diagrams in fig.~\ref{fig:diag_list} was evaluated within IR CHPT in ref.~\cite{beche2} and we have re-evaluated it 
independently. We keep the same labelling 
for the one-loop diagrams as in this reference for easier comparison. We agree with all the one-loop integrals 
given in detail there. Regarding their contributions to $A^{\pm}$ and $B^{\pm}$ 
we also agree with all of them except for the contributions of the so-called integral $I_B^{(2)}$, that results 
from the tensor one-loop integrals with one meson and two 
baryon propagators, see appendix C of ref.~\cite{beche2}. We find that systematically all 
its contributions as given in ref.~\cite{beche2} should be reversed in sign. These contributions appear in diagrams (c)+(d),
 (g)+(h) and (i). 
Apart from the direct calculation, we have checked that the expressions given in ref.~\cite{beche2} violate 
perturbative unitarity.  The latter  results because unitarity, eq.~\eqref{unita}, is a non-linear 
relation that mixes up orders in a power expansion. Denoting with a superscript the chiral order so that 
$T_{IJ\ell}=T_{IJ\ell}^{(1)}
+T_{IJ\ell}^{(2)}+T_{IJ\ell}^{(3)}+{\cal O}(q^4)$ the unitarity relation eq.~\eqref{unita} up to ${\cal O}(q^3)$ implies
\begin{align}
\hbox{Im}T_{IJ\ell}^{(3)}=\frac{|\vp|}{8\pi\sqrt{s}}\left(T_{IJ\ell}^{(1)}\right)^2~. 
\label{per_uni}
\end{align}

Our expressions fulfill eq.~\eqref{per_uni}, and those of ref.~\cite{beche2} also do once 
 the sign in front of $I_B^{(2)}$ is reversed for all its contributions.\footnote{The authors of ref.~\cite{beche2} state in page 30 that the scattering amplitude calculated obeys perturbative unitarity. 
It seems then that the difference in the sign referred above corresponds to a typo of \cite{beche2}.} Technically we follow the 
general procedure of refs.~\cite{becher} for calculating within IR and we do not give any expression 
for the different integrals calculated here because they were given already in refs.~\cite{kubis_vfm,beche2,becher}.

Now we proceed to compare our perturbative calculation with the experimental phase shifts for
 the low-energy data on the $\pi N$  $S$- and $P$-waves (which are the relevant partial waves for such energies.)
 Since our solution is perturbative one should evaluate 
 the phase shifts in a chiral expansion too. From the relation between $S_{IJ\ell}$ and $T_{IJ\ell}$, eq.~\eqref{s.def}, one has
\begin{align}
T_{IJ\ell}&=\frac{8\pi\sqrt{s}}{|\vp|}\sin \delta_{IJ\ell} \,e^{i\delta_{IJ\ell}}~,\nn\\
\cos\delta_{IJ\ell} \sin\delta_{IJ\ell}&=\frac{|\vp|}{8\pi\sqrt{s}}\hbox{Re}T_{IJ\ell}~.
\end{align}
Which implies that $\delta$ starts at ${\cal O}(q^2)$ so that up to ${\cal O}(q^4)$ one can write
\begin{align}
\delta_{IJ\ell}=\frac{|\vp|}{8\pi\sqrt{s}}\hbox{Re}T_{IJ\ell}~,
\label{delta.pert}
\end{align}
with $T_{IJ\ell}$ evaluated in the IR CHPT series  (in our present case up to ${\cal O}(q^3)$.) 

\begin{figure}[ht]
\psfrag{ss}{{\small $\sqrt{s}$ (GeV)}}
\psfrag{S11per}{$S_{11}$}
\psfrag{S31per}{$S_{31}$}
\psfrag{P11per}{$P_{11}$}
\psfrag{P13per}{$P_{13}$}
\psfrag{P31per}{$P_{31}$}
\psfrag{P33per}{$P_{33}$}
\centerline{\epsfig{file=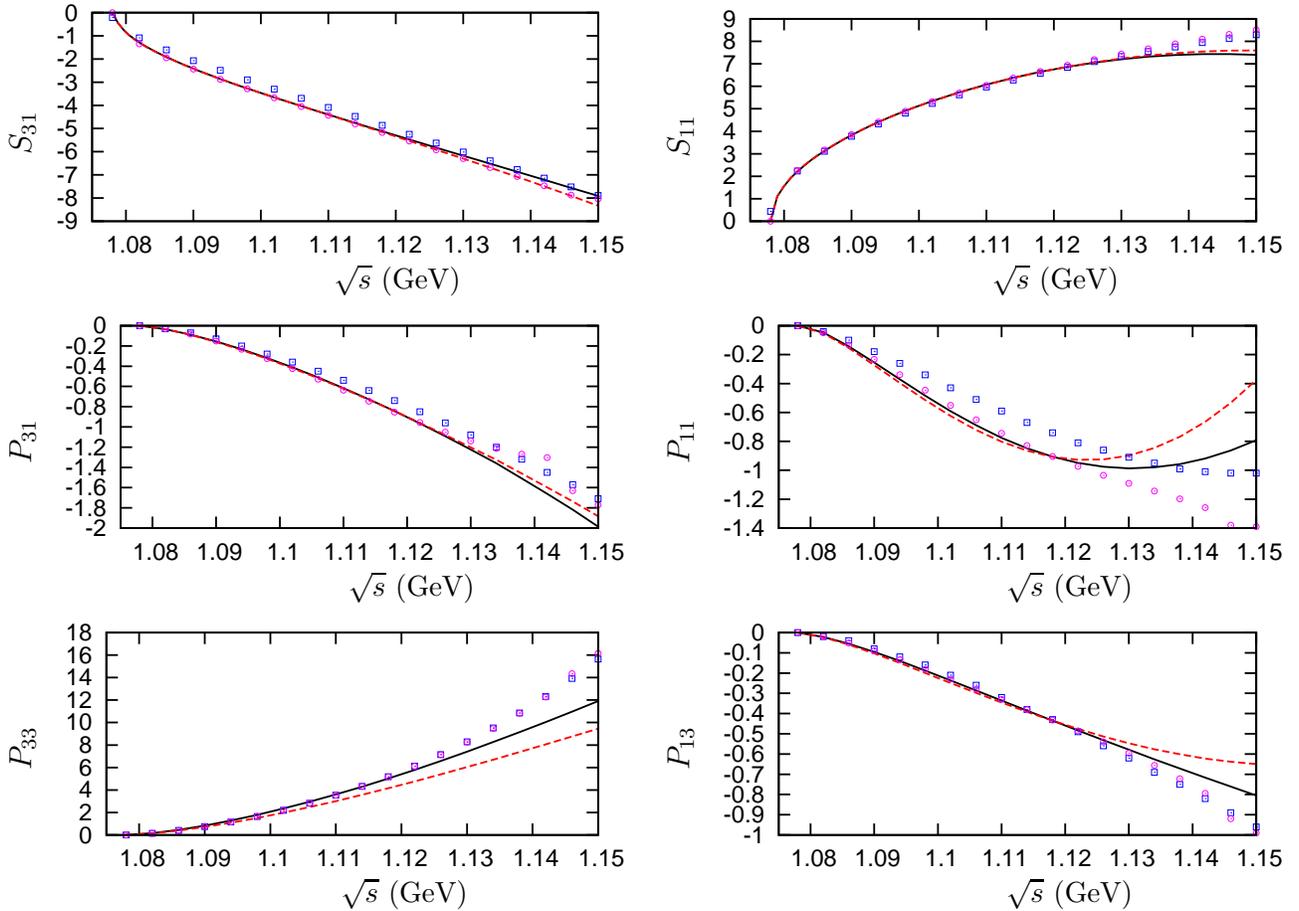,width=.7\textwidth,angle=-90}}
\vspace{0.2cm}
\caption[pilf]{\protect \small (Color online.) Fits to the KA85 pion-nucleon phase shifts \cite{ka84} as a 
function of $\sqrt{s}$ (in GeV) for $\sqrt{s}_{max}=1.13~$GeV in IR CHPT at ${\cal O}(q^3)$.  The KA85-1 fit corresponds to the 
solid curves and the KA85-2 fit to the dashed ones.  Data points: circles are KA85 and squares WI08 data. 
 \label{fig:res.ir.pert.ka85}}
\end{figure} 

We now consider the reproduction of the $\pi N$ phase shifts of the partial wave analyses of the Karlsruhe (KA85) group \cite{ka84}  and the current one  of
 the GWU  (WI08)  group \cite{wi08}.  
The fits are done with the full  IR CHPT calculation  to ${\cal O}(q^3)$. Due to the absence of error in these analyses \cite{ka84,wi08} there is some 
ambiguity in the definition of the $\chi^2$. Here we follow a similar strategy to that of ref.~\cite{pin} and define an error assigned to every point as the sum in quadrature of a systematic plus a statistical error,
\begin{align}
\hbox{err}(\delta)=\sqrt{e_s^2+e_r^2 \delta^2}~,
\label{err.def}
\end{align}
where $e_s$ is the systematic error and $e_r$ the relative one. In ref.~\cite{fettes3} a relative error of $3\%$ was taken while in ref.~\cite{pin} a 5\% error was considered. In the following we take for $e_s$ just 0.1 degrees and $e_r=2\%$. Regarding these values for the errors notice that isospin breaking corrections in $\pi N$ scattering are estimated to be rather small (ref.~\cite{fettes_com} 
estimates for $S$-waves an isospin breaking correction $\lesssim 1\%$.) We then consider the larger $2\%$ value as a safer estimate for isospin breaking effects not taken into account in our isospin symmetric study. Notice also that  the ${\cal O}(q^4)$ contributions are expected to be suppressed compared with the leading term by a relative 
 factor $\sim (M_\pi/\Lambda)^3\sim (0.14/0.5)^3\sim 0.02$. Although small, a finite value for $e_s$ helps to stabilize fits. Otherwise, 
 with $e_s=0$, extra weight is given to the small energy region close to threshold, where the phase shifts are smaller and absolute errors decrease. Tiny differences between the calculation and points in the input become then exceedingly relevant. We take $e_s=0.1$ degrees since it is much smaller than typical values of the phase shifts and  is also the typical size for the difference between the phase shifts of refs.~\cite{ka84,wi08} in the low-energy region for the $P_{11}$ partial wave (compare the squares \cite{ka84} and the circles \cite{wi08} in fig.~\ref{fig:res.ir.pert.ka85}.) We have also convinced ourselves that changes in these values for $e_s$ and $e_r$ do not affect our conclusions. 

The $\chi^2$ function to be minimized is defined in a standard way as
\begin{align}
\chi^2&=\sum_{i}\frac{(\delta-\delta_{th})^2}{\hbox{err}(\delta)^2}~,
\label{chi2.def}
\end{align}
with $\delta_{th}$ the phase shift calculated theoretically. For the minimization process we employ the program MINUIT \cite{cern}.

\begin{table}[ht]
 \begin{center}
\begin{tabular}{|r|r|r|r|r|r|r|}
\hline
{\small LEC}  & KA85-1        & KA85-2   & HBCHPT                            & HBCHPT       & HBCHPT                          & RS      \\
                 &                  &    & ${\cal O}(q^3)$  \cite{fettes3}  & Disp. \cite{buttiker}  & ${\cal O}(q^3)$  \cite{aspects} & \cite{aspects} \\
\hline
$c_1$            &  $-0.71\pm 0.49$ &  $-0.79\pm 0.51$   & $(-1.71,-1.07)$ &$-0.81\pm 0.12$& $-1.02\pm 0.06$  &     \\ 
$c_2$            &  $ 4.32 \pm 0.27$&  $3.49\pm 0.25$   & $(3.0,3.5)$     &$8.43\pm 56.9$ & $3.32\pm 0.03$   &3.9  \\ 
$c_3$            &  $-6.53 \pm 0.33$&  $-5.40\pm 0.13$ & $(-6.3,-5.8)$     &$-4.70\pm 1.16$& $-5.57\pm 0.05$  &$-5.3$ \\
$c_4$            &  $3.87\pm 0.15$  &  $3.32\pm 0.13$ & $(3.4,3.6)$       &$3.40\pm 0.04 $&                  &3.7  \\
\hline
$d_1+d_2$        &  $2.48\pm 0.59$  &  $0.94\pm 0.56$  & $(3.2,4.1)$      &&                  &     \\
$d_3$            & $-2.68 \pm 1.02$ &  $-1.10\pm 1.16$ & $(-4.3,-2.6)$     &&                  &     \\
$d_5$            & $2.69 \pm 2.20$  &  $1.86\pm 2.28$  & $(-1.1,0.4)$    &&                  &     \\
$d_{14}-d_{15}$  & $-1.71\pm 0.73$  &  $1.03\pm 0.71$& $(-5.1,-4.3)$      &&                  &     \\
$d_{18}$         & $-0.26\pm 0.40$  &  $-0.07\pm0.44$& $(-1.6,-0.5)$     &&                  &     \\
\hline
\end{tabular}
{\caption[pilf]{\protect \small Columns 2--5: Values of the low-energy constants for the KA85-1 and KA85-2 fits. The $c_i$ are given in
 GeV$^{-1}$ and the $d_i$ (or their combinations) in GeV$^{-2}$.  The renormalization scale for  $d_i(\lambda)$ is $\lambda=1$~GeV. The interval of values obtained in \cite{fettes3} by fitting low-energy $\pi N$ scattering data with HBCHPT at ${\cal O}(q^3)$ is given in the fourth column. Other determinations are given in columns fifth \cite{buttiker} and sixth \cite{aspects}. Resonance saturation estimates are collected in the last column \cite{aspects}. 
 \label{table.cs.ds.pert.ka85}}}
\end{center}
\end{table}

First, we discuss the reproduction of the KA85 data \cite{ka84} and later the  WI08 \cite{wi08} ones. As a first strategy, we fit directly these data from threshold up to an upper value denoted by $\sqrt{s}_{max}$, and consider several values for $\sqrt{s}_{max}$. A data point is included every 4~MeV in $\sqrt{s}$. One  observes that the $\chi^2$ per degree of freedom ($\chi^2_{d.o.f.}$) is below 1 for $\sqrt{s}_{max}\lesssim 1.13$~GeV, and then rises fast with energy so that for $\sqrt{s}_{max}=1.14$~GeV the $\chi^2_{d.o.f.}$ is 2.1 and for $\sqrt{s}_{max}=1.15~$GeV it becomes 3.6. In fig.~\ref{fig:res.ir.pert.ka85} we show by the solid line the result of the fit for $\sqrt{s}_{max}=1.13~$GeV.
At the level of the resulting curves the differences are small when varying  $\sqrt{s}_{max}$ within  the range indicated above. A good reproduction of the data is achieved up to around $\sqrt{s}\lesssim 1.14$~GeV, a similar range of energies  to that 
obtained in the ${\cal O}(q^3)$ HBCHPT fits of Fettes and Mei{\ss}ner \cite{fettes3}. From fig.~\ref{fig:res.ir.pert.ka85} one can readily see the origin of the rise in the $\chi^2$ with increasing $\sqrt{s}_{max}$.  It stems from the last points of the partial waves $P_{33}$, $P_{31}$ and $P_{11}$ for which the resulting curves depart from them, getting worse as the energy increases. The fast rising of the $P_{33}$ phase shifts is due to the $\Delta(1232)$ resonance. Though the tail of this resonance is mimicked in CHPT by the LECs, its energy dependence is too steep to be completely accounted for at ${\cal O}(q^3)$ due to the closeness of the $\Delta(1232)$ to the $\pi N$ threshold.  Indeed, this deficiency already occurred in the ${\cal O}(q^3)$ HBCHPT calculation of ref.~\cite{fettes3}. However, at ${\cal O}(q^4)$ the fit to data improves because of the appearance of new higher order LECs \cite{fettes4}. 

The resulting values for the CHPT LECs are shown in the second column of table~\ref{table.cs.ds.pert.ka85}, denoted by KA85-1, in units 
of GeV$^{-1}$ and GeV$^{-2}$ for the $c_i$ and $d_i$, respectively.  
 Note that at ${\cal O}(q^3)$ only the combinations of counterterms $d_1+d_2$, $d_3$, $d_5$, $d_{14}-d_{15}$ and $d_{18}$ appear in $\pi N$ scattering. The first four combinations were already explicitly shown in the expression for ${\cal L}_{\pi N}^{(3)}$, eq.~\eqref{lagN}. The counterterm $d_{16}$ does not appear because it is re-absorbed in the physical value of the pion-nucleon axial-vector coupling $g_A$,  once the lowest order $g$ 
constant is fixed in terms of the former \cite{beche2}. Under variations of  $\sqrt{s}_{max}$ most of the counterterms present a rather stable behavior, with  the  ${\cal O}(q^3)$ ones being the most sensitive.  The change in the LECs when varying $\sqrt{s}_{max}$ between 1.12 to 1.15~GeV is 
a source of uncertainty that is added in quadrature with the statistical error from the fit with $\sqrt{s}_{max}=1.13$~GeV, which has  a $\chi^2_{d.o.f.}$ of 0.9. The central values shown correspond to the same fit too. We also show in the table the values obtained from other approaches at ${\cal O}(q^3)$ \cite{aspects,mojzis,fettes3,buttiker}, including the 
${\cal O}(q^3)$ HBCHPT fit to $\pi N$ data \cite{fettes3}, the dispersive analysis within the Mandelstam triangle of ref.~\cite{buttiker} and the results at ${\cal O}(q^3)$ from ref.~\cite{aspects}, that also includes an estimation of the ${\cal O}(q^2)$ LECs from resonance saturation (RS).  Within uncertainties, our values for $c_1$, $c_3$ and $c_4$ are compatible with these other determinations. Instead, $c_2$  is somewhat 
larger, which is one of the main motivations for considering other fits to $\pi N$ scattering following the so called strategy 2, as explained below. Our values are also 
compatible with those determined from the $\pi N$ parameters up to ${\cal O}(q^4)$ in ref.~\cite{akaki_gasser} that gives the intervals 
$c_1=(-1.2, -0.9)$, $c_2=(2.6, 4.0)$ and $c_3=(-6.1, -4.4)$. The threshold parameters taken in this analysis are those calculated in 
ref.~\cite{ka84}. 
 Regarding the ${\cal O}(q^3)$ counterterms the comparison with HBCHPT is not so clear due to the large uncertainties both from our side as well as from \cite{fettes3}.  As discussed in more detail below, the ${\cal O}(q^3)$ contribution is typically the smallest between the different orders studied so that it is harder to pin down precise values for these counterterms. Indeed, we observe from the second column in table \ref{table.cs.ds.pert.ka85} that $d_3$, $d_5$ and $d_{14}-d_{15}$ have large errors, much larger than those of the ${\cal O}(q^2)$ counterterms (although the error estimated for $c_1$ is also large because 
the fits are not very sensitive to this counterterm which is multiplied by the small $M_\pi^2$ without energy dependence.) Our values for the LECs $d_i$, again within the large uncertainties, are compatible 
with those of ref.~\cite{fettes3}. Only $d_{14}-d_{15}$ is larger in our case, out of the range given in 
 \cite{fettes3} by around a factor 2. 

The threshold parameters for the fit KA85-1 are collected in the second column of table \ref{table.ir.pert.as.ka85}. We have evaluated the different scattering lengths and volumes by performing an effective range expansion (ERE) fit to our results in the low-energy region (namely, for  $|\vp|<M_\pi^2(1-M_\pi^2/m^2)$ which sets the range of the ERE.)\footnote{Numerical problems arising for $|\vp|\to 0$ prevent to calculate directly the threshold parameters as $\lim_{\vp\to 0} |\vp| \hbox{Re T}/8\pi\sqrt{s}|\vp|^{1+2 L}$.} The error given to our threshold parameters is 
just statistical. It is so small because the values of the scattering lengths and volumes are rather stable under changes of $\sqrt{s}_{max}$ and LECs within their uncertainties (taking into account the correlation among them.)  If treated in an uncorrelated way the error would be much larger. We also vary the numbers of terms in the ERE expansion from 3 to 5 and the slight variation in the resulting scattering lengths/volumes is also taken into account in the errors given.  
 In the last two columns of table \ref{table.ir.pert.as.ka85}, we give the values from the partial wave analyses of refs.~\cite{ka84,wi08}. Notice  that the differences between the central values from the latter two references are larger than 
one standard deviation, except for the $P_{33}$ case. 
 The differences between the $S_{31}$ scattering lengths and $P_{13}$ scattering volumes are specially large. Given this situation we consider that our calculated scattering lengths and volumes are consistent with the values obtained in the KA85 and WI08 partial wave analyses, except for the $P_{33}$ one for which our result is significantly larger. It is also too large compared with the values obtained in the ${\cal O}(q^3)$ HBCHPT fits to phase-shifts of ref.~\cite{fettes3}.

\begin{table}[ht]
 \begin{center}
\begin{tabular}{|r|r|r|r|r|r|r|}
\hline
{\small Partial} &   KA85-1       &    KA85-2           & KA85                 & WI08 \\
{\small Wave}     &              &                   &                      &      \\
\hline
$a_{S_{31}}$  & $ -0.100\pm 0.001$   & $-0.103\pm0.001$  &  $-0.100\pm 0.004$   & $-0.084$\\
$a_{S_{11}}$  & $ 0.171\pm 0.001$    & $0.172\pm 0.002$  &$0.175\pm 0.003$      & $0.171$\\
$a_{0+}^+$    &  $-0.010\pm 0.001$   & $-0.011\pm 0.001$ &$-0.008^a$            & $-0.0010\pm 0.0012$  \\
$a_{0+}^-$    &  $0.090\pm 0.001$    & $0.092\pm 0.001$  &  $0.092^a$           & $0.0883\pm 0.0005$  \\ 
$a_{P_{31}}$  &  $-0.052\pm 0.001$   & $-0.051\pm 0.001$ & $-0.044\pm 0.002$    &  $-0.038$ \\
$a_{P_{11}}$  &  $-0.078 \pm 0.001$  & $-0.088\pm 0.001$ & $-0.078\pm 0.002$    &  $-0.058$ \\
$a_{P_{33}}$  &  $0.251 \pm 0.002$   & $0.214 \pm 0002$  & $0.214\pm 0.002$     &  $0.194$ \\
$a_{P_{13}}$  &  $-0.034\pm 0.001$   & $-0.035\pm 0.001$ &$-0.030\pm 0.002$     &  $-0.023$ \\
\hline
\end{tabular}
{\caption[pilf]{\protect \small  $S$-wave scattering  lengths and $P$-wave scattering volumes in units of $M_\pi^{-1}$ and $M_\pi^{-3}$, respectively. Our results for the fits to the KA85-1 and KA85-2 are given in  the second and third columns, respectively. The fourth column corresponds to the values of the 
KA85 analysis \cite{ka84}. The values for WI08 are extracted from ref.~\cite{wi08} and the errors, when given, from ref.~\cite{fa02}.\\
$^a$ These numbers are given without errors because no errors are provided in ref.~\cite{ka84}. They are deduced from the KA85 ones for $a_{S_{31}}$ and $a_{S_{11}}$.    
\\
 \label{table.ir.pert.as.ka85}}}
\end{center}
\end{table}

Due to the large values for $c_2$ and $a_{P_{33}}$ we consider that the fit KA85-1 is not completely satisfactory and try a second  strategy (KA85-2). As  it was commented above, the rapid increase in the phase shifts due to the tail of the $\Delta(1232)$ is not well reproduced at ${\cal O}(q^3)$. As a result, instead of fitting the $P_{33}$ phase shifts as a function of energy we fit now the function $\tan \delta_{P_{33}}/|\vp|^3$ for three points with energy less than 1.09~GeV, where $\delta_{P_{33}}$ is the phase shifts for the $P_{33}$ partial wave. 
The form of this function is, of course, dictated by the ERE and at threshold it 
directly gives the corresponding scattering volume. We take a 2\% of error for these points because within errors this is the range of values spanned in table \ref{table.ir.pert.as.ka85} by the KA85 and WI08 results for $a_{P33}$. A relative error of 2\% was also taken for $e_r$ in eq.~\eqref{err.def}. The resulting values for the LECs are given in the third column of Table~\ref{table.cs.ds.pert.ka85} and the curves for $\sqrt{s}_{max}=1.13$~GeV are shown in 
fig.~\ref{fig:res.ir.pert.ka85} by the dashed lines, that have a  $\chi^2_{d.o.f.}=0.86$. We observe that these curves are quite similar to the ones previously obtained in KA85-1. Nevertheless, for the $P_{11}$ partial wave the description is slightly worse above 1.12~GeV and it is the main contribution to the final $\chi^2$. For the $P_{33}$ phase shifts one also observes a clear difference between the two curves as  the dashed line runs lower than the solid line. The former reproduces the standard values for the $P_{33}$ scattering volume, see column three of Table~\ref{table.ir.pert.as.ka85}, while for the latter it is larger. This is another confirmation that the description of the rapid rise of the $P_{33}$ phase shifts at ${\cal O}(q^3)$ enforces the fit to enlarge the value of the resulting scattering volume.  
  It is remarkable that now the value of the ${\cal O}(q^2)$ LEC $c_2$ is smaller and perfectly compatible with the interval of values of \cite{fettes3}. It is also interesting to note  that $c_3$ is also smaller, which is a welcome feature especially for two- and few-nucleon systems that are rather sensitive to large sub-leading two-pion exchange $NN$ potential that is generated by the inclusion of the $c_1$, $c_3$ and $c_4$ \cite{kaiser_peri}. See refs.~\cite{epe1,epe2,fews} for a thorough discussion on this issue for two- and few-nucleon systems.   Related to this point, one has determinations of $c_3$ and $c_4$ by a partial wave analysis of the $pp$ and $np$ scattering data 
 from ref.~\cite{rent_cs} with the results
\begin{align}
c_3&=-4.78\pm 0.10~\hbox{GeV}^{-1}~,\nn\\
c_4&=+3.96\pm 0.22~\hbox{GeV}^{-1}~.
\end{align}
The systematic errors are not properly accounted for yet in these determinations due to the dependence on the matching point that distinguishes  between the long-range part of the $NN$ potential (parameterized from CHPT) and the short-range one (with a purely phenomenological parameterization.) Namely, the same authors in ref.~\cite{remt_prl} considered this issue and when varying the matching point from 1.8~fm to 1.4~fm the LECs changed significantly: $c_3=-5.08(28)\to -4.99(21)$ and $c_4=4.70(70)\to 5.62(69)$~GeV$^{-1}$. 
 With respect to the ${\cal O}(q^3)$ counterterms we see that the central values have shifted considerably compared with KA85-1. This clearly indicates that these LECs cannot be properly pinned down by fitting $\pi N$ scattering data. Within uncertainties $d_3$, $d_5$ and $d_{18}$ overlap at the level of one sigma. The LECs $d_{1}+d_2$ and $d_{14}-d_{15}$ require to take into account a variation of 2 sigmas. In view of this situation we consider that one should be conservative  and give  ranges of values for these latter combination of LECs in order to make them compatible 
\begin{align}
d_1+d_2&=+0.4\,\ldots \,+3~\hbox{GeV}^{-2}~,\nn\\
d_{14}-d_{15}&=-2.4\,\ldots\, +1.75~\hbox{GeV}^{-2}~.
\label{dbad.ka85}
\end{align}
These values correspond to the minimum and maximum of those shown in the second and third columns of table~\ref{table.cs.ds.pert.ka85} allowing  a variation of one sigma.

 The scattering lengths and volumes for KA85-2 are collected in the third column of Table~\ref{table.ir.pert.as.ka85}. They are calculated 
from our results similarly as explained above for  the KA85-1 fit.
It is remarkable that now  the value for the $P_{33}$ scattering volume is perfectly compatible with the determinations from KA85 and WI08.
 We see a good agreement between our ${\cal O}(q^3)$ IR CHPT results and the scattering lengths/volumes for KA85. Only the $P_{11}$ scattering volume is slightly different, though  the difference between the KA85 and WI08 results is significantly large for this case too. 
One also observes differences beyond the error estimated in KA85 for the $P_{13}$ scattering volume between the KA85 and WI08 values. Ours is closer to the KA85 one. 

It is also worth emphasizing that our fits to the phase shifts  of the KA85 analysis (KA85-1 and KA85-2), as shown in fig.~\ref{fig:res.ir.pert.ka85}, offer a good reproduction of the data and the worsening for higher energies stems in a smooth way as in ${\cal O}(q^3)$ HBCHPT \cite{fettes3}. This is certainly an improvement compared with the previous $\pi N$ study in IR CHPT to ${\cal O}(q^3)$ of ref.~\cite{elli2}. In this latter reference, data could only be fitted up to around 1.12~GeV and large discrepancies above that energy, rapidly increasing with energy,  emerged in the $S_{31}$, $P_{13}$ and $P_{11}$ partial waves.

\begin{figure}[ht]
\psfrag{ss}{{\small $\sqrt{s}$ (GeV)}}
\psfrag{S11per}{$S_{11}$}
\psfrag{S31per}{$S_{31}$}
\psfrag{P11per}{$P_{11}$}
\psfrag{P13per}{$P_{13}$}
\psfrag{P31per}{$P_{31}$}
\psfrag{P33per}{$P_{33}$}
\centerline{\epsfig{file=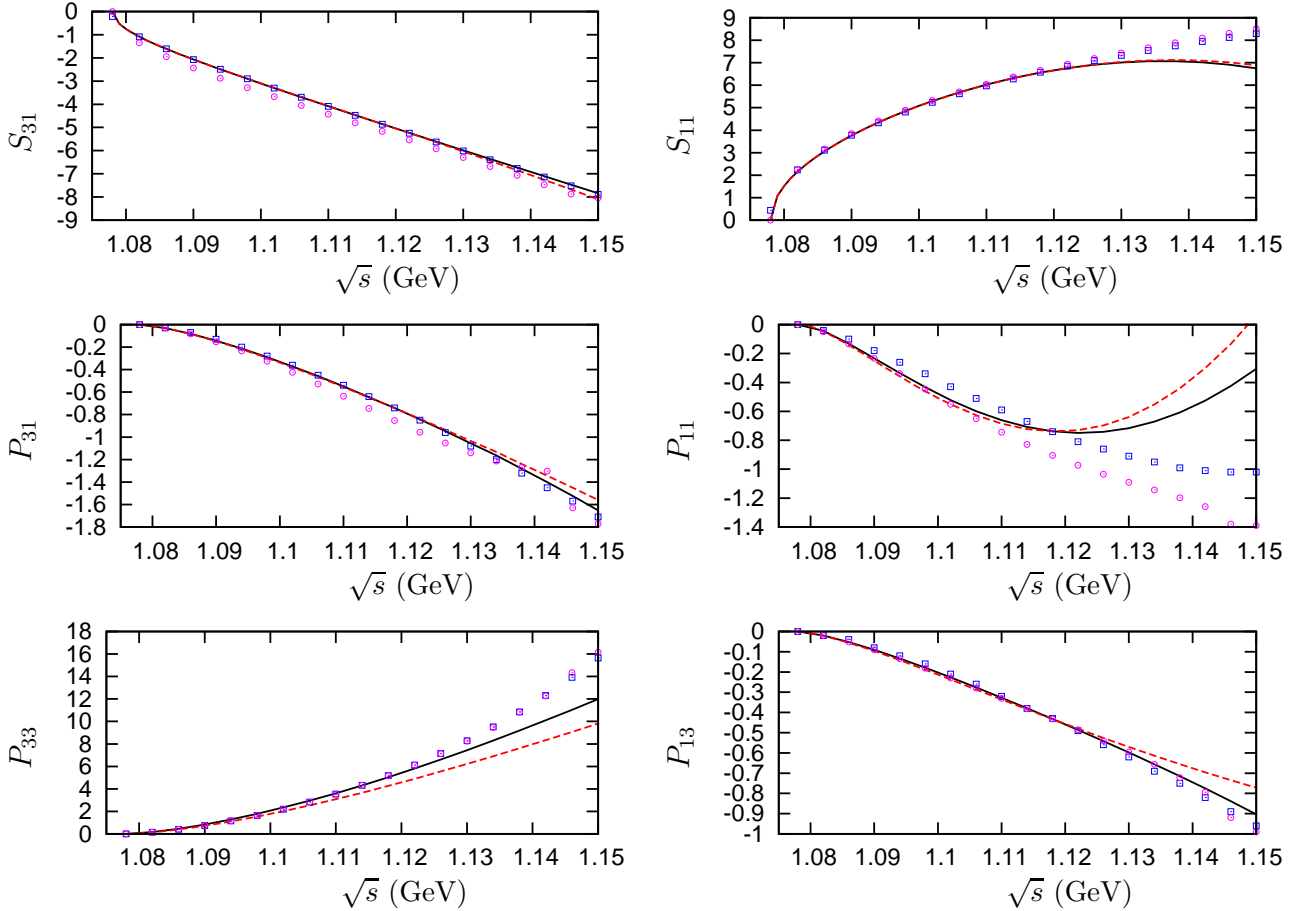,width=.7\textwidth,angle=-90}}
\vspace{0.2cm}
\caption[pilf]{\protect \small (Color online.)  Fits to the WI08 pion-nucleon phase shifts \cite{wi08} as a 
function of $\sqrt{s}$ (in GeV) for $\sqrt{s}_{max}=1.13~$GeV in IR CHPT at ${\cal O}(q^3)$.  The WI08-1 fit corresponds to the 
solid curves and the WI08-2 fit to the dashed ones.  Data points: circles are KA85 and squares WI08 data.
 \label{fig:res.ir.pert.wi08}}
\end{figure} 

We proceed along similar lines and perform fits of type 1 and 2 to the current solution of the GWU group \cite{wi08} (WI08). These fits are denoted by WI08-1 and WI08-2, in that order. The resulting curves for $\sqrt{s}_{max}=1.13~$GeV are shown by the solid and dashed lines in fig.~\ref{fig:res.ir.pert.wi08}, respectively. One observes very similar curves to the KA85-1 and KA85-2 fits except for the $P_{11}$ phase shifts. Here, the agreement with the WI08 data is considerably worse. This has a clear translation into the values of the $\chi^2$ for this partial wave,  which increases almost by a factor 3,  from 20 (KA85-2) to 55 (WI08-2) (the number of fitted points is 12.) It is clear from fig.~\ref{fig:res.ir.pert.wi08} that IR CHPT at ${\cal O}(q^3)$ does not compare well with the $P_{11}$ WI08 phase shifts even at  very low energies, $\sqrt{s}<1.11$~GeV. For the KA85 data the situation is much better, compare with fig.~\ref{fig:res.ir.pert.ka85}. Indeed, previous solutions of the GWU (and prior VPI) group had a behavior similar to that of KA85 for the $P_{11}$ phase-shifts,  e.g. the solution SM01 employed in the analysis of ref.~\cite{elli2} also using IR CHPT at ${\cal O}(q^3)$. In view of the difficulties of our study based on 
IR CHPT at  ${\cal O}(q^3)$ for reproducing the $P_{11}$ phase shifts of WI08 at low energies we consider advisable 
a revision of the current solution WI08 of the GWU group and 
the way the $\eta N$ data affect the low-energy $P_{11}$ phase shifts in the coupled channel approach followed \cite{briscoe}. 
 
The other distinctive features when comparing strategies 1 and 2 for the fits to the WI08 data are similar to those already discussed for the KA85  fits. In this way, one has for WI08-1 that $c_2$ and $c_3$ have a value in modulus  larger by around 1~GeV$^{-1}$ than for WI08-2. 
Related to this,  the $P_{33}$ scattering volume is also significantly larger for WI08-1 than for WI08-2. The values of the fitted LECs for WI08-1 and WI08-2  are collected in the second and third columns of Table~\ref{table.ir.pert.cs.ds.wi08}.   
 One observes that the resulting LECs at ${\cal O}(q^2)$ are quite similar between KA85-1, WI08-1, on the one hand, and KA85-2, WI08-2, on the other, so that within uncertainties they are compatible in either of the two strategies. 
In the third column of Table~\ref{table.ir.pert.cs.ds.wi08} we present the average of the LECs from our fits in Tables~\ref{table.cs.ds.pert.ka85} and \ref{table.ir.pert.cs.ds.wi08}. 
The error given for every LEC is the sum in quadrature of the largest of the statistical errors shown in the previous tables and the one resulting from the dispersion in the central values. This is a conservative procedure which recognizes that both strategies are acceptable for studying low-energy $\pi N$ scattering and that takes into account the dispersion in the LECs that results from 
 changes in the data set.
 Within errors, the values of the LECs in the last column of Table~\ref{table.ir.pert.cs.ds.wi08} are compatible 
with those from  HBCHPT at ${\cal O}(q^3)$ ($d_{14}-d_{15}$ is the only counterterm that differs by more than one standard deviation 
from the interval of values of ref.~\cite{fettes3}.) 
 
 Regarding the threshold parameters we list in the second and third columns of Table~\ref{table.ir.pert.as.wi08} the values for the $S$-wave scattering lengths and $P$-wave scattering volumes corresponding to the fits WI08-1 and WI08-2 with $\sqrt{s}_{max}=1.13$~GeV (the same fits  shown in fig.~\ref{fig:res.ir.pert.wi08}.) The procedure for their determination is the same as the one already discussed for KA85-1. The largest changes compared with the values of the KA85-1 and KA85-2 fits, respectively, occur for the $a_{S_{31}}$ and $a_{P_{11}}$ scattering length and volume, in order. The latter also shows the largest difference between the results of the fits following strategy 1 and 2 (a 12\% of relative difference for the KA85 case and a 9\% for the WI08 one.) Nonetheless, neither of our results for $a_{P_{11}}$, including strategy 1 and 2 KA85 and WI08 fits, is compatible with the value of WI08 \cite{wi08}, shown in the last column of Table~\ref{table.ir.pert.as.ka85}. The largest difference occurs for the value of KA85-2 which is a 50\% smaller than the one from WI08. 
Coming back to the WI08 fits, we note that the isoscalar $S$-wave scattering length $a_{0+}^+$ is now a vanishing positive number while the $P_{11}$ scattering volume has decreased, and is compatible with the KA85 result within one sigma. We notice that the tiny errors estimated for the threshold parameters  resulting from our fits in Tables~\ref{table.ir.pert.as.ka85} and \ref{table.ir.pert.as.wi08} are just statistical and are determined in the same way as explained above for the KA85-1 fit.  Of course, systematic errors due to higher orders in the chiral expansion     and different data sets taken induce larger systematic uncertainties than the small errors shown. In this sense, the difference between the values obtained for each partial wave in these columns provides a better estimation of uncertainties. We then calculate the average\footnote{Not the weighted average. The given errors  are calculated by adding in quadrature for each LEC the largest of the errors in Tables~\ref{table.cs.ds.pert.ka85} and \ref{table.ir.pert.cs.ds.wi08} and the one resulting from the average of values.} of the four values for each scattering length/volume shown altogether in Tables~\ref{table.ir.pert.as.ka85} and \ref{table.ir.pert.as.wi08}.  This is given in the last column of Table~\ref{table.ir.pert.as.wi08}. We also see that our averaged values for the $a_{0+}^+$ and $a_{0+}^-$ scattering lengths are compatible with the results obtained in ref.~\cite{raha}, $a_{0+}^+=0.0015\pm 0.0022$ and $a_{0+}^-=0.0852\pm 0.0018$ $M_\pi^{-1}$, that takes into account isospin breaking corrections in the analysis of  recent experimental results on pionic hydrogen and pionic deuterium data

\begin{table}[ht]
 \begin{center}
\begin{tabular}{|r|r|r|r|}
\hline
LEC &   WI08-1       &   WI08-2  & Average \\
\hline
$c_1$           & $-0.27\pm 0.51$  & $-0.30\pm 0.48$  &  $-0.52\pm 0.60$\\
$c_2$           & $4.28\pm 0.27$    & $3.55\pm 0.30$  &  $3.91\pm 0.54$ \\
$c_3$           & $-6.76\pm 0.27$   & $-5.77\pm 0.29$ &  $-6.12\pm 0.72$\\
$c_4$           & $4.08\pm 0.13$    & $3.60\pm 0.16$  &  $3.72 \pm 0.37$ \\
$d_1+d_2$       & $2.53\pm 0.60$    & $1.16\pm 0.65$  &  $1.78\pm 1.1$\\
$d_3$           & $-3.65\pm 1.01$   & $-2.32\pm 1.04$ &  $-2.44\pm 1.6$\\
$d_5$           & $5.38\pm 2.40$    & $4.83\pm 2.18$  &  $3.69\pm 2.93$\\
$d_{14}-d_{15}$ & $-1.17\pm 1.00$   & $1.27\pm1.11$   &  $-0.145\pm 1.88$ \\
    $d_{18}$        & $-0.86\pm 0.43$   & $-0.72\pm 0.40$ &  $-0.48\pm 0.58$\\ 
\hline
\end{tabular}
{\caption[pilf]{\protect \small Fitted LECs in units of GeV$^{-1}$ ($c_i$) and GeV$^{-2}$ ($d_i$) for the fits WI08-1 and WI08-2 with $\sqrt{s}_{max}=1.13$~GeV. The last columns 
is the average of all the fits in Tables~\ref{table.cs.ds.pert.ka85} and \ref{table.ir.pert.cs.ds.wi08}.
 \label{table.ir.pert.cs.ds.wi08}}}
\end{center}
\end{table}

\begin{table}[ht]
 \begin{center}
\begin{tabular}{|r|r|r|r|}
\hline
Partial & WI08-1 & WI08-2 & Average \\
Wave   &         &        &          \\
\hline
  $a_{S_{31}}$ & $ -0.081\pm 0.001$    & $-0.082 \pm 0.001$ & $-0.092\pm 0.012$ \\
 $a_{S_{11}}$  & $ 0.165\pm 0.002$     & $0.167 \pm 0.002$  & $0.169\pm 0.004$  \\
 $a_{0+}^+$    &  $0.001\pm 0.001$     & $0.001 \pm 0.001$  & $-0.005\pm 0.007$ \\
 $a_{0+}^-$    &  $0.082\pm 0.001$     & $0.083 \pm 0.001$  & $0.087\pm 0.005$  \\ 
 $a_{P_{31}}$  &  $-0.048\pm 0.001$    & $-0.051 \pm 0.001$ & $-0.051\pm 0.002$ \\
 $a_{P_{11}}$  &  $-0.073\pm 0.001$    & $-0.080 \pm 0.001$ & $-0.080\pm 0.006$ \\
 $a_{P_{33}}$  &  $0.252\pm 0.002$     & $0.222 \pm 0.002$  & $0.232\pm 0.017$  \\
 $a_{P_{13}}$  &  $-0.032\pm 0.001 $   & $-0.035 \pm 0.001$ & $-0.034\pm 0.002$ \\
\hline
\end{tabular}
{\caption[pilf]{\protect \small  $S$-wave scattering  lengths and $P$-wave scattering volumes in units of $M_\pi^{-1}$ and $M_\pi^{-3}$, respectively, 
for the fits WI08-1 and WI08-2 with $\sqrt{s}_{max}=1.13$~GeV. The last column corresponds to the averaged values of the threshold parameters 
 of all the fits in Tables~\ref{table.ir.pert.as.ka85} and \ref{table.ir.pert.as.wi08}. 
 \label{table.ir.pert.as.wi08}}}
\end{center}
\end{table}

\begin{figure}[ht]
\psfrag{ss}{{\small $\sqrt{s}$ (GeV)}}
\psfrag{S11}{$S_{11}$}
\psfrag{S31}{$S_{31}$}
\psfrag{P11}{$P_{11}$}
\psfrag{P13}{$P_{13}$}
\psfrag{P31}{$P_{31}$}
\psfrag{P33}{$P_{33}$}
\centerline{\epsfig{file=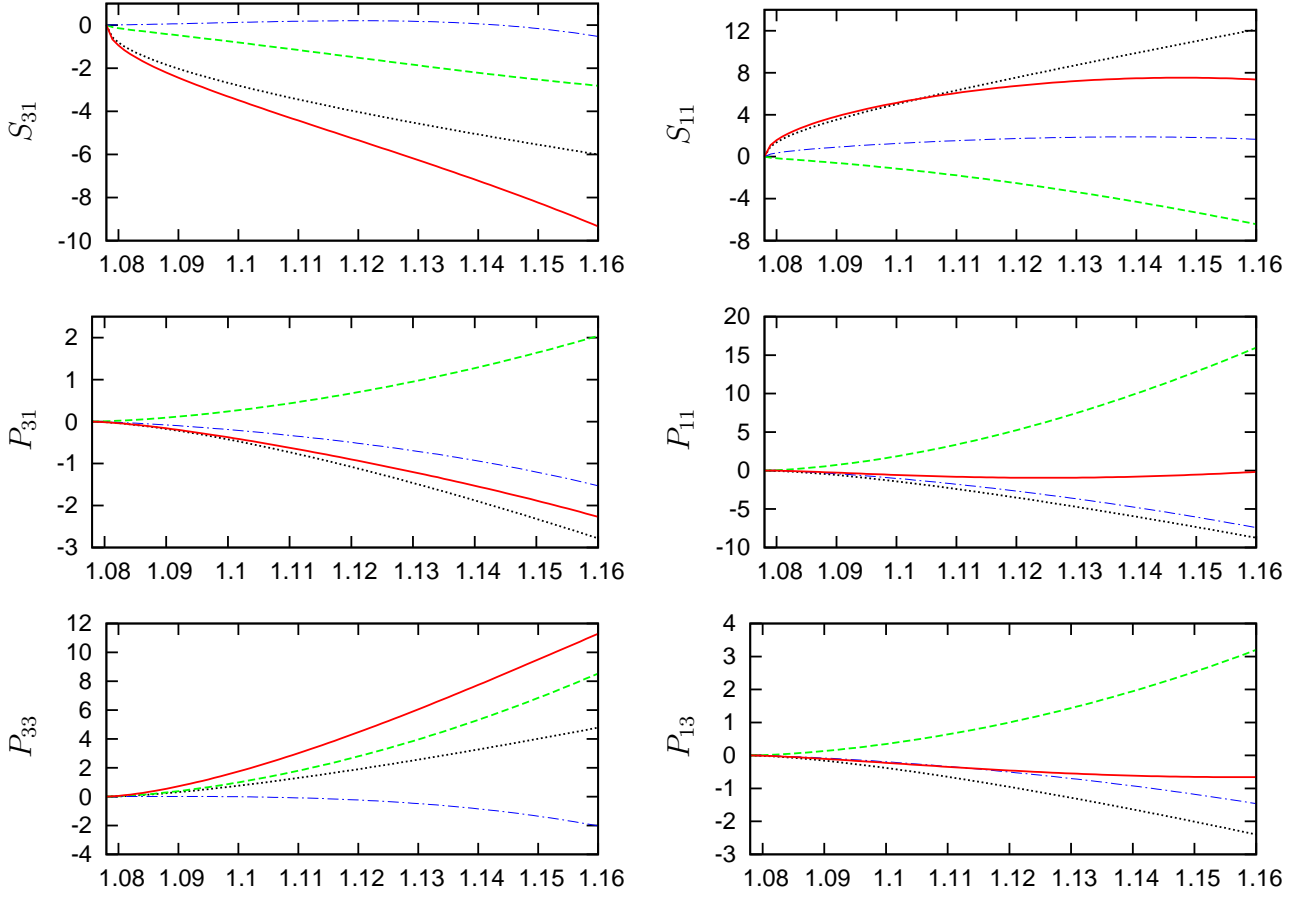,width=.7\textwidth,angle=-90}}
\vspace{0.2cm}
\caption[pilf]{\protect \small (Color online.) Different chiral orders contributing to the phase shifts for the KA85-1 fit. 
 The (black) dotted, (green) dashed and (blue) dash-dotted are the first, second and third order, respectively. The (red) solid 
line is the sum of all of them. 
 \label{fig:ordenes}}
\end{figure}

Finally, we show in fig.~\ref{fig:ordenes} the different chiral order contributions to the total phase shifts (depicted by the solid lines) for the fit KA85-1 (shown in fig.~\ref{fig:res.ir.pert.ka85} by the solid lines.) The dotted lines correspond to the leading result, the dashed ones to NLO and the dash-dotted ones to N$^2$LO. A general trend observed  is the partial cancellation between the ${\cal O}(q^2)$ and ${\cal O}(q^3)$ contributions. For the $P$-waves, the cancellation is almost exact at low energies while at higher energies the ${\cal O}(q^2)$ contribution is larger in modulus than the ${\cal O}(q^3)$ one (except for the $P_{31}$ partial wave where the cancellation is almost exact all over the energy range shown, so that the first order describes well this partial wave.) For the $S$-waves at low energies ($\sqrt{s}\lesssim 1.11$~GeV) the first order contributions dominates, though the second order one tends to increase rapidly with energy. For these partial waves the second order contribution is much larger than the third order one and the partial cancellation between these orders is weak (even both orders add with the same sign for $S_{31}$ at the highest energies shown.) The smallness of the third order contribution for the $S$-waves together with the fact that it is also clearly smaller than the second order one for most of the $P$-waves explain the difficulties to pin down precise values for the ${\cal O}(q^3)$ LECs (the $d_i$'s), as already indicated above. 

The LEC $d_{18}$ is important as it is  directly involved in the violation of the GT relation \cite{goldberger}. 
Up to ${\cal O}(M_\pi^3)$ one has \cite{fettes3,beche2}
\begin{align}
g_{\pi N}&=\frac{g_A m}{F_\pi}\left(1-\frac{2 M_\pi^2 d_{18}}{g_A}\right)~.
\label{goldberger}
\end{align}
We quantify the deviation from the GT relation by
\begin{align}
\Delta_{GT}&=\frac{g_{\pi N}F_\pi}{g_A m}-1~.
\label{delta.def}
\end{align}
Inserting our averaged value of $d_{18}$  in the third column of Table~\ref{table.ir.pert.cs.ds.wi08} into eq.~\eqref{goldberger}, 
we  then find 
\begin{align}
\Delta_{GT}&=0.015\pm 0.018~,
\label{gt.per}
\end{align}
which is compatible with the values around 2--3\% that are nowadays preferred 
from $\pi N$ and $NN$ partial wave analyses \cite{arndtcc,schroder,rentcc}. 
 In terms of the $\pi N$ coupling constant, from eq.~\eqref{goldberger} our value for $d_{18}$ translates in 
\begin{align}
g_{\pi N}&=13.07\pm 0.23
\end{align}
or $f^2=\left(g_{\pi N} M_\pi/4m\right)^2/\pi=0.077\pm 0.003$. Within uncertainties our result at strict ${\cal O}(M_\pi^3)$  
is compatible at the level of one sigma with the determinations of refs.~\cite{arndtcc,schroder,rentcc}.  

 However, IR CHPT at ${\cal O}(q^3)$ gives rise to a caveat concerning the GT relation.
 The point is that the full calculation at this order (IR CHPT contains higher orders due to the $1/m$ relativistic resummation) 
 produces  a huge GT relation violation of about a 20\%, similarly as in ref.~\cite{elli2}. For the evaluation of the GT relation discrepancy  
 in our present calculations we study the  $\pi^-p\to \pi^- p$ scattering. We select this particular 
process in the charge basis of states because the crossed $u$-channel process, $\pi^+p\to \pi^+p$, 
is purely $I=3/2$ and thus there is no  $u$-channel nucleon pole, which  requires the same quantum numbers as for the nucleon, in the isospin limit.
 Otherwise  the $s$- and $u$-channel nucleon poles 
overlap  for some values of the scattering angle.  When projecting the $u$-channel nucleon pole in a partial wave it produces a cut 
for $m^2-2M_\pi^2+M_\pi^4/m^2<s<m^2+2M_\pi^2$, with the branch points very close to the nucleon pole at $s=m^2$. As a result, there is not soft way to 
calculate the residue at the $s$-channel nucleon pole unless the $u$-channel nucleon pole is removed, as done by considering the $\pi^-p\to \pi^-p$ scattering. 
 The latter is finally projected in the partial wave $P_{11}$, with the same quantum numbers as the nucleon. 
The  ratio of the residues
 at the nucleon pole of the full ${\cal O}(q^3)$ IR CHPT partial wave and the direct ($s$-channel) Born term calculated with $g_A$, $M_\pi$ and $m$ at
 their physical values, gives us directly the ratio between the squares of the full pion-nucleon coupling and the one from the GT relation.\footnote{Note that there is no crossed Born term for $\pi^- p\to \pi^- p$ and that  
the LO Born term in term of physical parameters satisfies exactly the GT relation.} 
Numerically we find that the full calculation gives rise to a violation of the GT relation of around 20-25\%, while its strict ${\cal O}(M_\pi^3)$ restriction 
is much smaller, eq.~\eqref{gt.per}. Related to this one has a significant renormalization scale dependence on the GT violation.\footnote{Eq.~\eqref{gt.per} is   renormalization scale independent because  the beta function for $d_{18}$ is zero \cite{fettes3}.} In this way, for the fit KA85-1 (second column of Table~\ref{table.cs.ds.pert.ka85}) at $\lambda=1$~GeV one has a 22\% of violation of the GT relation while for $\lambda=0.5$~GeV a 15\% stems. On the other hand,  ref.~\cite{gasser2} performed a relativistic calculation of $\Delta_{GT}$ directly in dimensional regularization within the $\overline{MS}-1$ renormalization scheme and obtained 
a natural (much smaller) and renormalization scale independent loop contribution to $\Delta_{GT}$. It seems then  that the problem that we find for the calculation of $\Delta_{GT}$ with IR, obtained earlier in ref.~\cite{elli2}, is related to the peculiar way the chiral counting is restored in the IR approach \cite{pinto,gorgorito}.  
 We tentatively conclude that  a neat advance in the field would occur once a relativistic regularization method were available that conserved the chiral counting in the evaluation of loops while, at least, avoided any residual renormalization scale dependence.

\section{Unitarized amplitudes and higher energies}
\label{sec4}
\def\theequation{\arabic{section}.\arabic{equation}}
\setcounter{equation}{0}

In order to resum the right-hand cut or unitarity cut we consider the unitarization method of refs.~\cite{pin,plb}, to which we refer
for further details. Notice that this method does not only provide a unitary $\pi N$ amplitude but also takes care of the 
analyticity properties associated with the right-hand cut. In ref.~\cite{pin} this approach was used for unitarizing 
the ${\cal O}(q^ 3)$ HBCHPT $\pi N$ partial waves from ref.~\cite{fettes3}. However, no explicit Lorentz-invariant one-loop 
calculation for $\pi N$ scattering has been 
unitarized in the literature yet. This is an interesting point since by taking explicitly into account the presence of the 
unitarity cut the rest of the amplitude is expected to have a softer chiral expansion. According to ref.~\cite{plb} we 
express ${\cal T}_{IJ\ell}$ as 
\begin{align}
T_{IJ\ell}&=\frac{1}{{\cal T}_{IJ\ell} ^{-1}+g(s)}~,
\label{basic}
\end{align}
where the unitarity pion-nucleon loop function is given by
\begin{align}
g(s)&=\frac{1}{(4\pi)^2}\biggl\{
a_1+\log\frac{m^2}{\mu^2}-\frac{M_\pi^2-m^2+s}{2s}\log\frac{m^2}{M_\pi^2}
+\frac{|\vp|}{\sqrt{s}}\biggl[\log(s-\Delta+2\sqrt{s}|\vp|)\nn\\
&+\log(s+\Delta+2\sqrt{s}|\vp|)
-\log(-s+\Delta+2\sqrt{s}|\vp|)-\log(-s-\Delta+2\sqrt{s}|\vp|)
\biggr]\biggr\}\,,
\label{g.def}
\end{align}
with   $\Delta=M_{\pi}^2-m^2$. The interaction kernel ${\cal T}_{IJ\ell}$ has no right-hand cut and is determined by 
matching order by order with the perturbative chiral expansion of $T_{IJ\ell}$ calculated in CHPT.  In this way, with $g(s)=
{\cal O}(q)$, one has \cite{plb}
\begin{align}
T_{IJ\ell}^{(1)}+T_{IJ\ell}^{(2)}+T_{IJ\ell}^{(3)}={\cal T}_{IJ\ell}^{(1)}+{\cal T}_{IJ\ell}^{(2)}
+{\cal T}_{IJ\ell}^{(3)}
-g(s)\left(T_{IJ\ell}^{(1)}\right)^2~,
\end{align}
so that
\begin{align}
{\cal T}_{IJ\ell}^{(1)}&=T_{IJ\ell}^{(1)}~,\nn\\
{\cal T}_{IJ\ell}^{(2)}&=T_{IJ\ell}^{(2)}~,\nn\\
{\cal T}_{IJ\ell}^{(3)}&=T_{IJ\ell}^{(3)}+g(s)\left(T_{IJ\ell}^{(1)}\right)^2~,
\label{det}
\end{align}
and ${\cal T}_{IJ\ell}={\cal T}_{IJ\ell}^{(1)}+{\cal T}_{IJ\ell}^{(2)}+{\cal T}_{IJ\ell}^{(3)}$ is then replaced in eq.~\eqref{basic}. 
Since the resulting partial wave is now unitary, we calculate the phase shifts directly from the relation 
$T_{IJ\ell}= \frac{8\pi\sqrt{s}}{|\vp|}e^{i\delta_{IJ\ell}}\sin \delta_{IJ\ell}$ that follows from 
eqs.~\eqref{s.def} and \eqref{s.def.2}.

The subtraction constant $a_1$ is determined by requiring that $g(s)$ vanishes at the nucleon mass $s=m^2$. In this way the $P_{11}$ partial-wave 
has the nucleon pole at its right position, otherwise it would disappear. This is due to the fact that for this partial wave 
${\cal T}_{\frac{1}{2}\frac{1}{2}1}^{-1}$ vanishes at $s=m^2$ so it is required that $g(m^2)=0$. Otherwise  $T_{\frac{1}{2}\frac{1}{2}1}$, 
eq.~\eqref{basic}, would be finite at $s=m^2$. 

\begin{figure}[ht]
\psfrag{ss}{{\small $\sqrt{s}$ (GeV)}}
\psfrag{S11per}{$S_{11}$}
\psfrag{S31per}{$S_{31}$}
\psfrag{P11per}{$P_{11}$}
\psfrag{P13per}{$P_{13}$}
\psfrag{P31per}{$P_{31}$}
\psfrag{P33per}{$P_{33}$}
\centerline{\epsfig{file=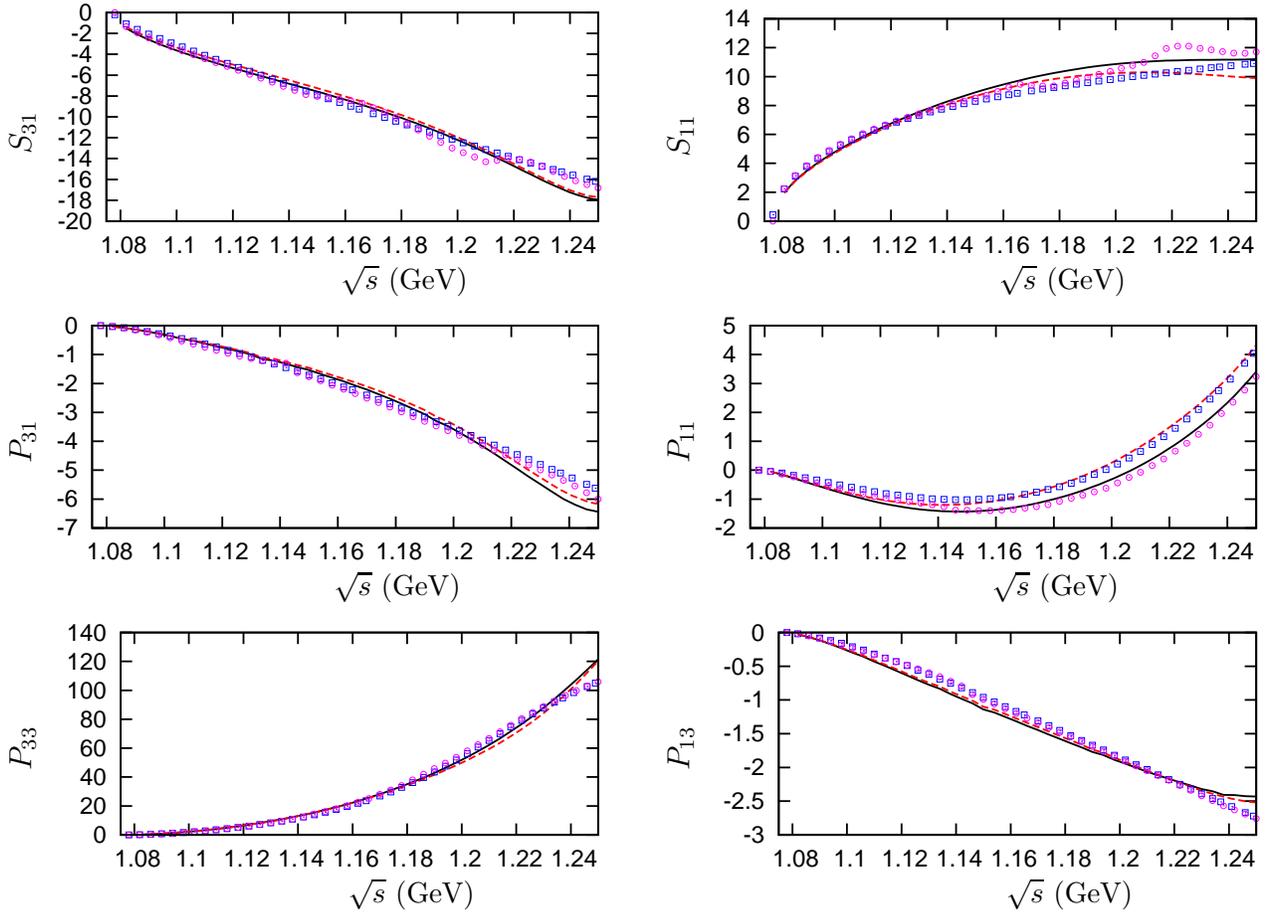,width=.7\textwidth,angle=-90}}
\vspace{0.2cm}
\caption[pilf]{\protect \small (Color online.)  Fits to the KA85 and WI08 pion-nucleon phase shifts as a 
function of $\sqrt{s}$ (in GeV) employing the unitarized $\pi N$ amplitudes, eq.~\eqref{uni.cdd}. 
 The solid (dashed) lines correspond to the fit of the KA85  (WI08) data. 
 \label{fig:ir.uni}}
\end{figure}

Due to the closeness of the $\Delta(1232)$ resonance to the $\pi N$ threshold it is expedient to implement  a method to  take into account its presence in order to provide a higher energy description of $\pi N$ phase-shifts beyond the purely perturbative results discussed in section~\ref{sec3}. 
 As commented in the introduction we can add a CDD pole \cite{cdd} in the $P_{33}$  channel so as to reach the region of the $\Delta(1232)$ 
resonance. The addition of the CDD pole conserves the discontinuities of the partial wave amplitude across the cuts. A CDD pole corresponds to 
a zero of the partial wave-amplitude along the real axis and hence to a pole in the inverse of the amplitude. We then modify eq.~\eqref{basic} 
by including such a pole in $T_{\frac{3}{2}\frac{3}{2}1}^{-1}$, 
\begin{align}
T_{\frac{3}{2}\frac{3}{2}1}=\Biggl({\cal T}_{\frac{3}{2}\frac{3}{2}1}^{-1}+\frac{\gamma}{s-s_P}+g(s)\Biggr)^{-1}~,
\label{uni.cdd}
\end{align}
where $\gamma$ and $s_P$ are the residue and pole position of the CDD pole, in order, so that two new free parameters 
 enter. The amplitude  ${\cal T}_{IJ\ell}$ is determined as in eq.~\eqref{det}. We also distinguish here between the fits to the KA85 \cite{ka84} and WI08 \cite{wi08} phase-shifts. The fits are done up to $\sqrt{s}=\sqrt{s}_{max}=1.25$~GeV for all the partial waves. One cannot afford to go to higher energies because of an intrinsic limitation of IR CHPT. Additional unphysical cuts and poles are generated by the infinite order resummation of the sub-leading $1/m$ kinetic energy terms accomplished in IR \cite{pinto,gorgorito,bernard}. In our case the limiting circumstance is the appearance of a pole when the Mandelstam variable $u=0$.\footnote{Many of the tensor integrals involved in the one-loop calculations of $\pi N$ scattering develop such a pole. In particular, it arises in the simplest  scalar two-point loop function $I(u)$, following the notation of ref.~\cite{becher}.} When projecting in the different partial waves this singularity gives rise to a strong branch point at $s=2(m^2+M_\pi^2)\simeq 1.34^2~$GeV$^2$, which indicates the onset of a non-physical right-hand cut that extends to infinity and that produces strong violation of unitarity. This translates into strong rises of the phase-shifts calculated employing eq.~\eqref{uni.cdd} for energies $\sqrt{s}\gtrsim 1.26$~GeV. This is why we have taken $\sqrt{s}_{max}=1.25$~GeV because for higher 
energies these effects are clearly visible in the calculated phase-shifts. The $\chi^2$ to be minimized is the same as already used for the pure perturbative study, eq.~\eqref{chi2.def}, employing also the same definition for err$(\delta)$. The resulting fits are shown in fig.~\ref{fig:ir.uni}, where the solid lines correspond to the fit of the KA85 data and the dashed ones to WI08. One can see a rather good agreement with data in the whole energy range from threshold up to 1.25~GeV, including the reproduction of 
the raise in the $P_{33}$ phase shifts associated with the $\Delta(1232)$ resonance. 
 The improvement is manifest in the $P_{11}$ partial wave although some discrepancy with the WI08 
data in the lower energy region remains, 
 being better the agreement with KA85 phase-shifts. 
Compared with the perturbative treatment of section~\ref{sec3}  one observes a drastic increase in the range of energies for which a globally acceptable description of the data is achieved.

The values of the resulting LECs are collected in Table~\ref{table.cs.ds.uni}. We consider that the pure perturbative study of section~\ref{sec3} is the proper way to determine the chiral LECs. The new values in Table~\ref{table.cs.ds.uni} do not constitute an alternative determination to those offered in Tables~\ref{table.cs.ds.pert.ka85} and \ref{table.ir.pert.cs.ds.wi08} and should be employed within UCHPT studies.  Nonetheless, it is remarkable that the values for the LECs obtained are compatible with the average of values given in the fourth column of Table~\ref{table.ir.pert.cs.ds.wi08}, in particular, for the ${\cal O}(q^2)$ LECs the central values are also rather close to the fitted values in Table~\ref{table.cs.ds.uni}. 
 Since we have a procedure to generate the $\Delta(1232)$ resonance through the CDD pole in eq.~\eqref{uni.cdd}, such agreement is  surprising  since the contribution of this resonance to the LECs is very important \cite{aspects}. The point is that the typical value of $\gamma/(s-s_P)$ in the low-energy region studied in section~\ref{sec3} is only around a factor 2 larger in modulus than the subtraction constant $a_1/(4\pi)^2$ in eq.~\eqref{g.def}, being the latter a quantity of first chiral order. As a result, at low energies, the CDD pole gives a contribution that can be computed as ${\cal O}(q^3)$, since the lowest order ones comes from 
$-( {T^{(1)}_{IJL}})^2 \gamma/(s-s_P)$. This explains why the values of the second order LECs are preserved, despite having included the CDD pole. 

\begin{table}[ht]
 \begin{center}
\begin{tabular}{|r|r|r|r|r|r|}
\hline
LEC &   Fit &   Fit   & Partial & Fit  &  Fit  \\
     &	KA85 &   WI08  & Wave    & KA85 &  WI08 \\
\hline
$c_1$           & $-0.48\pm 0.51$   & $-0.53\pm 0.48$     & $a_{S_{31}}$ & $-0.115$ &  $-0.104$ \\
$c_2$           & $4.62\pm 0.27$    & $4.73\pm 0.30$      & $a_{S_{11}}$ & $0.152$ &  $0.150$ \\
$c_3$           & $-6.16\pm 0.27$   & $-6.41\pm 0.29$     & $a_{0+}^+$   & $-0.026$ &  $-0.020$ \\
$c_4$           & $3.68\pm 0.13$    & $3.81\pm 0.16$      & $a_{0+}^-$   & $0.089$ &   $0.085$\\
$d_1+d_2$       & $2.55\pm 0.60$    & $2.70\pm 0.65$      & $a_{P_{31}}$ & $-0.050$ &  $-0.048$ \\
$d_3$           & $-1.61\pm 1.01$   & $-1.73\pm 1.04$     & $a_{P_{11}}$ & $-0.080$ & $-0.075$ \\
$d_5$           & $0.93\pm 2.40$    & $1.13\pm 2.18$      & $a_{P_{33}}$ & $0.245$ &  $0.250$\\
$d_{14}-d_{15}$ & $-0.46\pm 1.00$   & $-0.61\pm1.11$       & $a_{P_{13}}$ & $-0.41$ &  $-0.039$ \\
    $d_{18}$        & $0.01\pm 0.21$   & $-0.03\pm 0.20$ &              &  &  \\ 
\hline
\end{tabular}
{\caption[pilf]{\protect \small Fitted LECs in units GeV$^{-1}$ ($c_i$) and GeV$^{-2}$ ($d_i$)
 for the fits KA85 and WI08 employing the unitarized partial waves. 
 We also give the scattering lengths and volumes in units of $M_\pi$ and $M_\pi^{-3}$, respectively. 
 \label{table.cs.ds.uni}}}
\end{center}
\end{table}

The values of the resulting threshold parameters with the present unitarized amplitudes are collected in the last two columns 
of Table~\ref{table.cs.ds.uni}. We observe that all of them are compatible with the averaged values given in the last column 
of Table~\ref{table.ir.pert.as.wi08}.  The $P_{33}$ scattering volume turns out a bit too high in the lines 
of the values obtained with the perturbative fits following strategy 1, despite the reproduction of the $\Delta(1232)$ resonance. 
 Finally, we also mention that similarly huge values for the GT violation are also obtained  from the unitarized amplitudes as in the 
pure perturbative treatment.  Indeed, the same value for $\Delta_{GT}$, eq.~\eqref{delta.def}, is obtained in the unitarized case for the same values of the
 LECs because $g(m^2)=0$ (there is no CDD pole in the $P_{11}$ partial wave.)

\section{Summary and conclusions}
\label{sec5}

We  studied  elastic pion-nucleon scattering employing covariant CHPT up-to-and-including ${\cal O}(q^3)$ in Infrared Regularization \cite{becher}.  We  followed two strategies for fitting the phase shifts provided the partial wave 
analysis of refs.~\cite{ka84,wi08}. In one of them, instead of fitting the $P_{33}$ phase-shifts, we  considered the reproduction 
of the function $|\vp|^3/\tan \delta_{P_{33}}$ around the threshold region (for $\sqrt{s}\leq 1.09$~GeV.) The rational behind this is 
to reduce the impact of the $\Delta(1232)$ when performing fits to data, 
avoiding the rapid rise of phase-shifts with energy that tends to increase the value of the resulting scattering volume. An accurate reproduction of  
pion-nucleon phase-shifts up to around 1.14~GeV results. The main difference between both strategies has to do with the values of the ${\cal O}(q^2)$ LECs $c_2$ and $c_3$, that are smaller in absolute value for strategy 2 fits. As expected, the $P_{33}$ scattering volume is also smaller for these  fits and compatible with previous determinations. We have  discussed separately the fits to  data of the Karlsruhe \cite{ka84} and GWU \cite{wi08} groups. We obtain a much better reproduction of the $P_{11}$ phase shifts for the former partial wave analysis. IR     CHPT at ${\cal O}(q^3)$ is not able to reproduce the $P_{11}$ phase shifts of the current solution of the GWU group even at very low energies. This suggests that a revision of this solution would be in order.  The averaged values for the LECs and threshold parameters resulting from the two strategies and all data sets are given in the last columns of Tables~\ref{table.ir.pert.as.wi08} and \ref{table.ir.pert.cs.ds.wi08} in good agreement with other previous determinations. The reproduction of experimental phase-shifts  is similar in quality to that obtained previously with ${\cal O}(q^3)$ HBCHPT \cite{fettes3}, showing also a smooth onset of  the departure from experimental data  for higher energies.  This is an improvement compared with previous work \cite{elli2}. 
 In addition, we obtain a small violation of the 
Goldberger-Treiman relation at strict ${\cal O}(M_\pi^3)$,  compatible with present determinations.
However, the deviation from the Goldberger-Treiman relation is still a caveat because when  all the terms in the full IR CHPT calculation 
at ${\cal O}(q^3)$ are kept the resulting discrepancy is much higher, around 20-30\%. 

We have also employed the non-perturbative methods of Unitary CHPT \cite{plb,pin} to resum the right-hand cut of the pion-nucleon partial waves.  The $\Delta(1232)$ resonance is incorporated in the approach 
as a Castillejo-Dalitz-Dyson pole in the inverse of the amplitude. A good reproduction of the phase shifts is reached  for $\sqrt{s}$ up to around 1.25~GeV. There is an intrinsic limitation in IR CHPT for reaching higher energies due to the presence of a branch cut at  $s=2(m^2+M_\pi^2)\simeq 1.34^2~$GeV$^2$. Above that energy strong violations of unitarity occurs due to the onset of an unphysical cut associated with the infinite resummation of relativistic corrections accomplished in IR. This also originates a strong rise of phase-shifts noticeable already for $\sqrt{s}\gtrsim 1.25$~GeV. The values of the LECs at ${\cal O}(q^2)$ is compatible to those obtained with the pure perturbative study.

\section*{Acknowledgements}
We thank  R.~Workman for useful correspondence in connection with GWU group partial-wave analyses and 
SAID program.
  This work is partially funded by the grants MEC  FPA2007-6277, FPA2010-17806 and the Fundaci\'on S\'eneca 11871/PI/09.
 We also thank the financial support from  the BMBF grant 06BN411, the EU-Research Infrastructure
Integrating Activity
 ``Study of Strongly Interacting Matter" (HadronPhysics2, grant n. 227431)
under the Seventh Framework Program of EU and   
the Consolider-Ingenio 2010 Programme CPAN (CSD2007-00042). JMC acknowledges  the MEC contract FIS2006-03438, the EU Integrated Infrastructure Initiative Hadron Physics Project contract RII3-CT-2004-506078 and the  Science and Technology 
Facilities Council [grant number ST/H004661/1] for support.


  \end{document}